\documentclass[prd,showpacs,showkeys,nofootinbib,12pt,amsfonts]{revtex4}
\usepackage{epsfig}
\usepackage{amssymb,amsthm,graphicx,citesort}
\usepackage[ansinew]{inputenc}
\usepackage{latexsym}
\usepackage{stmaryrd}
\usepackage{multibox}
\usepackage{graphicx, epic, eepic, color}
\usepackage{dcolumn}
\usepackage{bm}
\usepackage{epsfig}
\usepackage[ansinew]{inputenc}
\def\czslash{{\nearrow\!\!\!\!\!\!\!{\cal Z}}}
\def\riz{\nearrow\!\!\!\!\!\!\!{\it Riz}}

\def\lapl{ {\,  {\circ}\!-\!\!\! -{\bullet}\,}  }
\def\qslash{\not\!Q}
\def\zslash{{\not\!Z}}
\def\a{\alpha}
\def\b{\beta}
\def\g{\gamma}
\def\vt{\vartheta}
\def\d{\delta}
\def\vta{\vartheta}
\def\oz{{\overline{\zeta}}}
\def\sqr#1#2{{\vcenter{\hrule height.#2pt\hbox{\vrule width.#2pt
height#1pt \kern#1pt \vrule width.#2pt}\hrule height.#2pt}}}
\def\square{\mathchoice\sqr64\sqr64\sqr{4.2}3\sqr{3.0}3}

\def\negenspace{\kern-1.1em}
\def\qslash{\nearrow\!\!\!\!\!\!\!Q}

\def\zslash{\nearrow\!\!\!\!\!\!\!Z}

\def\zslash{\nearrow\!\!\!\!\!\!\!Z}
\def\Xislash{\nearrow\!\!\!\!\!\!\!{\Xi^{~\alpha\beta}}}

\def\Riz{\nearrow\!\!\!\!\!\!{{\rm Riz}}}
\def\cRiz{\nearrow\!\!\!\!\!\!{{\cal R}iz}}

\def\negenspace{\kern-1.1em}

\def\negenspaceexp{\kern-0.5em}

\def\a{\alpha}
\def\b{\beta}
\def\g{\gamma}
\def\c{\gamma}

\def\d{\delta}

\def\l{\lambda}

\def\o{\omega}

\def\pa{\partial}

\def\ie{{\it i.e., }}

\def\cl{{\cal L}}

\def\cz{{\cal Z}}
\begin{document}
\title{Linear connections with propagating spin-3 field in gravity}

\author{\footnotesize PETER
  BAEKLER\footnote{peter.baekler@fh-duesseldorf.de
}}

\affiliation{Fachbereich Medien, Fachhochschule D\"usseldorf\\ University
  of Applied Sciences\\ Josef-Gockeln-Str.\ 9\\40474 D\"usseldorf,
  Germany
}
\author{{\footnotesize NICOLAS BOULANGER\footnote{nicolas.boulanger@umh.ac.be}}}
\affiliation{Universit\'e de Mons-Hainaut, Acad\'emie
Wallonie-Bruxelles\\
M\'ecanique et Gravitation, Avenue du Champ de Mars 6\\ B-7000
Mons, Belgium}

\author{{\footnotesize FRIEDRICH W.\ HEHL}\footnote{hehl@thp.uni-koeln.de
}}

\affiliation{Institute for Theoretical Physics\\ University of Cologne\\ 50923
  K\"oln, Germany\\ and\\ Department of Physics and Astronomy\\
  University of Missouri-Columbia\\ Columbia, MO 65211, USA}


\begin{abstract}
We show that Fronsdal's Lagrangian for a free massless spin-3 gauge
  field in Minkowski spacetime is contained in a general
  Yang--Mills-like Lagrangian of metric-affine gravity (MAG), the
  gauge theory of the general affine group in the presence of a
  metric.  Due to the geometric character of MAG, this can best be
  seen by using Vasiliev's frame formalism for higher-spin gauge
  fields in which the spin-3 frame is identified with the {\it
    tracefree nonmetricity} one-form associated with the shear
  generators of $GL(n,\mathbb{R})$.  Furthermore, for specific
  gravitational gauge models in the framework of full nonlinear MAG,
  exact solutions are constructed, featuring propagating massless and
  massive spin-3 fields.
\end{abstract}



\maketitle


\section{Introduction}

Metric-affine gravity (MAG, see \cite{PRs} for a review) constitutes a
rich and natural framework for the study of gravitational phenomena at
high-energy, when spacetime is expected to lose its Riemannian
character. Thanks to its geometric formulation, it is also a promising
candidate theory for the unification of gravity with the other
fundamental forces based on Yang--Mills-like actions for internal gauge
groups.  Its spacetime can be seen as a generalization of the
spacetime of Weyl's unified theory of gravitation and electromagnetism
\cite{Weyl:1918}. There, spacetime is described by a manifold in which
not only the direction but also the norm of vectors are affected by
parallel-transport, thereby providing a ``true infinitesimal
geometry''.  By adopting a metric-affine spacetime $(L_n,g)$, instead
of the usual Riemannian spacetime $V_n$ of Einstein's general
relativity, one naturally extends the latter by introducing torsion
$T$ and nonmetricity $Q$, besides the Levi-Civita connection, still
conserving a classical, smooth spacetime.  
The connection one-form $\Gamma$ in $(L_n,g)$ takes value in the Lie algebra
$gl(n,\mathbb{R})$ of the general linear group $GL(n,\mathbb{R})$, 
subgroup of the affine gauge group of MAG. 
More precisely, a metric-affine spacetime is described by a metric 
$g_{\alpha\beta}$, a coframe field ${\vartheta}^{\alpha}$ and an independent 
connection ${\Gamma}_{\alpha}{}^{\beta}$ that generally carries torsion
$T^\a:=D\vartheta^\a$ and nonmetricity $Q_{\alpha\beta}:=-Dg_{\alpha\beta}$, 
where $D$ denotes the $GL(n,\mathbb{R})$-covariant exterior derivative. 

The idea that the metricity condition $Q_{\alpha\beta}=0$ may become
operational at low energy after spontaneous symmetry breaking is
attractive and has been investigated for some time (see \cite{PRs} and
references therein).  That the totally symmetric piece of the
nonmetricity may become massive after the spontaneous symmetry
breaking of $GL(n,\mathbb{R})$ down to its Lorentz subgroup
$SO(1,n-1)$, leaving the metric as massless Goldstone field, was
studied recently in \cite{Kirsch:2005st}.  There, it was suggested
that this totally symmetric and traceless piece of $Q$ should behave
as a massless spin-3 gauge field at the Planck energy.

It is well known that the nonmetricity $Q$ contains a spin-3 piece,
the totally symmetric and traceless piece of $Q$ being called {\tt
  trinom} in \cite{PRs}.  However, it is only in the recent work
\cite{Boulanger:2006tg} that this idea was taken seriously: Fronsdal's
action \cite{Fronsdal:1978rb} for a massless spin-3 field was written
such that on-shell the propagating spin-3 field coincides with {\tt
  trinom.}  The latter field then acquired mass by a specific
Brout--Englert--Higgs (BEH) mechanism based on the spontaneous breaking
$GL(n,\mathbb{R})/SO(1,n-1)$ viewed as a small part of a more general
BEH mechanism by which the full diffeomorphism group $G=Di\!f\!f(n,R)$
is broken down to its Lorentz subgroup $H$.  Besides the metric being
regarded as a Goldstone field, specific parameters characterizing the
coset space $G/H$ were interpreted \cite{Boulanger:2006tg} as higher-spin 
connections, in the context of which it seemed plausible indeed
to assume the Lorentz group as stability group $H$.

In the present work, we elaborate on the idea that the totally
symmetric and traceless part of the nonmetricity could represent a
massless spin-3 gauge field. 
After a brief review of MAG geometry in Section \ref{sec:MAG}, 
we show in Section \ref{Vasiliev} that MAG houses indeed such a
field by exhibiting Fronsdal's theory for a massless spin-3 
field as a subsector of linearized MAG.  Due to the
geometric nature of MAG, it is actually more convenient to consider
Vasiliev's Lagrangian \cite{Vasiliev:1980as} for a massless frame-like 
spin-3 field in Minkowski spacetime. 
The crucial step is to identify the traceless nonmetricity --- 
the component of the nonmetricity which lies along the shear 
generator of $GL(n,\mathbb{R})$ --- with
Vasiliev's spin-3 frame-like field, thereby providing another geometrical
interpretation for the latter field and showing that Fronsdal's spin-3
theory is hidden in MAG.

To take care of this observation, in Section \ref{sec:exsol}
different types of field Lagrangians
of MAG and the corresponding field equations will be investigated in
the sector of vanishing torsion.  Our main aim is to show that the
field equations of MAG provide solutions for propagating spin-3
fields.

As a first ansatz we present in (\ref{ansatz_Q1}) (subsection \ref{ANSATZ}) 
a nonmetricity $Q_{\alpha\beta}$ that is pretty much adapted to
describe propagating modes of the totally symmetric spin-3 field
$^{(1)}Q_{(\alpha\beta\gamma )}$.  Furthermore, we investigate
particular Lagrangians to exhibit the different propagation behavior
of massless as well as of massive modes.

In subsection \ref{PURE1Q} we consider a Yang--Mills-like Lagrangian
for the pure spin-3 field $^{(1)}Q_{\alpha\beta\gamma }$ and show
that, in vacuum, this field configuration is just trivial since from a
field theoretical point of view kinetic terms of the nonmetricity are
missing. This reminds of the situation in the Einstein--Cartan theory
where torsion is proportional to the spin of matter, just mediating
some type of contact interaction.

The situation can be improved in subsection \ref{+HilbertE} by adding 
curvature dependent terms to the Lagrangian (\ref{lag_Q1}). 
Adding a Hilbert--Einstein type Lagrangian supports the existence of 
massless spin-3 modes.  Because
of the curvature $\sim DQ_{\alpha}{}^{\beta}$ and the second field
equation $\sim DR_{\alpha}{}^{\beta}$, the Bianchi identities will
``freeze" out the genuine dynamical degrees of freedom of the fields.
Hence, a field Lagrangian such as (\ref{lag_RQ1}) leads still to a
second field equation which is algebraic in the field strengths, cf.\
(\ref{second_2}). In that case we would like to call such fields {\it
  pseudo}-propagating. Provided the coupling constants will be
adjusted suitably, the second field equation will be fulfilled without
further constraints and the first field equation reduces to an
Einstein equation with cosmological constant in a Riemannian
spacetime.

If we supplement the Lagrangian (\ref{lag_RQ1}) with further pieces of
the nonmetricity, cf. subsection \ref{allQ2}, additionally massive modes 
can be generated, at least for particular choices of the coupling constants.

After identifying the Vasiliev field $e_{\alpha\beta\gamma}$ with the
tracefree nonmetricity ${\qslash}_{\alpha\beta\gamma}$ of MAG, it is
natural to consider field Lagrangians quadratic in the
strain-curvature $Z_{\alpha\beta}$ yielding genuine dynamical degrees
of freedom, cf. subsection \ref{YM}. 
For this particular consideration a slightly modified
Kerr--Schild ansatz for the nonmetricity will be considered in which
the propagation will be characterized by the field ${\ell}$, cf.
(\ref{ansatz_Q1}).  Consequently, this type of approach will convert
the nonlinear second field equation into a {\em linear} partial
differential equation of second order. Accordingly, we derive $\square
\, ^{(1)}Q_{(\alpha\beta\gamma )}=0$ for the components of the spin-3
field for massless modes (${\ell}^2=0$).  Observe that in general
relativity this method implies that the full nonlinear Einstein tensor
equals to its linearized part. In this sense the Kerr--Schild ansatz
leads to an ``exact linearization", cf.\ G\"urses et al.
\cite{GurGur}. This linearizing property of the Kerr--Schild ansatz can
also be applied successfully in MAG.
We generalize the Kerr--Schild form ${\ell}$ in (\ref{ell2_nonzero}).
Then also field configurations with {\em massive} spin-3 character can
be generated. An example of such a simple toy-model can be found in
subsection \ref{YMmassive}. 

The conclusions are outlined in 
Section \ref{sec:discu} and some technical results are relegated to the 
appendices. 
%
\section{Metric-affine geometry}
\label{sec:MAG}

\subsection{Notation and conventions}
%
In this section we will summarize shortly the main properties of an
$n$-dimensional metric-affine spacetime. At each point of spacetime,
we have a coframe $\vartheta^\a$ spanning the cotangent space; the
frame (or anholonomic) indices $\a,\b,\g...$ run over $0,1,...,n-1$.
We denote local coordinates by $x^i$; (holonomic) coordinate indices
are $i,j,k,...=0,1,...,n-1$. Most of our formalism is correct for
arbitrary $n$. However, in this article we will mainly concentrate on
$n=4\,$.  We can decompose the coframe with respect to a coordinate
coframe according to $\vartheta^\a=e_i{}^\a\,dx^i$.  For the frame
$e_\a$, spanning the tangent space, we have
$e_\a=e^i{}_\a\,\partial_i$. If $\rfloor$ denotes the interior
product, then we have the duality condition $e_\a\rfloor
\vartheta^\b=\delta_\a^\b$.  Symmetrization will be denoted by
parentheses $(\a\b):=\frac 12\a\b+\frac 12\b\a$, antisymmetrization by
brackets $[\a\b]:= \frac 12\a\b-\frac 12\b\a$, and analogously for $p$
indices with the factor $\frac{1}{p!}$, see Schouten
\cite{Schouten54}. Indices excluded {}from (anti)symmetrization are
surrounded by vertical strokes: $(\a|\g|\b):=\frac 12\a\g\b+\frac 12
\b\g\a$, etc.

We assume the existence of a metric
\begin{equation}
  g = g_{\alpha\beta}\,{\vartheta}^{\alpha}\otimes\,
  {\vartheta}^{\beta}
\quad {\rm with}  \quad
  g_{ij}=e_i{}^\a e_j{}^\b\,g_{\a\b}\,.
\end{equation}
Choosing {\it orthonormal} (co)frames
$e_i^{~\a}\stackrel{*}{=}{\stackrel{\circ}{e}}_i{}^{\a}$, we have the
condition
\begin{eqnarray}
  g_{\a\b}\stackrel{*}{=}g_{ij}\,{\stackrel{\circ}{e}}{}^i{}_{\a}\,
{\stackrel{\circ}{e}}{}^j{}_{\b}
  =o_{\a\b}:={\rm diag}(-1,+1,\ldots,+1)\,,
\end{eqnarray}
whereas the {\it holonomic} gauge is defined by
$C^\a:=d\vt^\a\stackrel{*}{=}0$, that is,
\begin{equation}
\vartheta^\a\stackrel{*}{=}\delta_i^\a\,dx^i\,,\quad
e_\a\stackrel{*}{=}\delta^i_\a\,\partial_i\,.
\end{equation}

When a metric is present, we can introduce the Hodge star operator
$^\star$. If we denote exterior products of the coframe $\vt^\a$ as
$\vta^{\a\b}:=\vta^\a\wedge\vta^\b$,
$\vta^{\a\b\g}:=\vta^\a\wedge\vta^\b\wedge\vta^\g$, etc., then we can
introduce, as an alternative to the theta-basis, the eta-basis
according to 
\begin{equation}\label{etabasis}\eta:=\,^\star 1\,,\quad
  \eta^\a:=\,^\star\vta^\a\,,\quad\eta^{\a\b}:=\,^\star\vta^{\a\b}\,,
  \quad\eta^{\a\b\g}:=\,^\star\vta^{\a\b\g}\,,\; {\rm etc.,}
\end{equation}
see \cite{PRs,Thirring}. This basis can be very convenient if the
$^\star$ is involved in formulas.

Furthermore, the manifold will be assumed to carry a
metric-independent linear connection ${\Gamma}_{\alpha}{}^{\beta}$,
see Kobayashi \& Nomizu \cite{Koba} or Frankel \cite{Frankel}, that
generally supports the torsion
$T^{\alpha}:=D{\vartheta}^{\alpha}=d{\vartheta}^{\alpha}
+{\Gamma}_{\b}{}^{\alpha}\wedge {\vartheta}^{\b}$ and the
nonmetricity ${Q_{\alpha\beta}:=-Dg_{\alpha\beta}}$.  Here $d$ denotes
the exterior derivative and $D$ the $GL(n,\mathbb{R})$ gauge-covariant
exterior derivative.

It is of advantage to split the connection into Riemannian and
non-Riemannian parts.  If we introduce the distortion 1-form
$N_{\alpha}{}^{\beta}$, the connection reads
\begin{equation}\label{Gamma_N}
{\Gamma}_{\alpha}{}^{\beta} =
{\widetilde{\Gamma}}_{\alpha}{}^{\beta} + N_{\alpha}{}^{\beta}\,.
\end{equation}
In the following, the tilde denotes always the purely
Riemannian contribution. Torsion and nonmetricity can be recovered
{}from $N_{\alpha}{}^{\beta}$ by
\begin{equation}\label{dist_1}
Q_{\alpha\beta}=2N_{(\alpha\beta )}\, \quad {\rm and}\, \quad
T^{\alpha}=N_{\b}{}^{\alpha}\wedge {\vartheta}^{\b}\, .
\end{equation}
Explicitly, the distortion 1-form $N_{\alpha}{}^{\beta}$ can be
expressed in terms of torsion and nonmetricity as
\begin{equation}
  N_{\alpha\beta}=-e_{[\alpha}\rfloor T_{\beta]} + {1\over
    2}(e_{\alpha}\rfloor e_{\beta}\rfloor
  T_{\gamma})\,\vartheta^{\gamma} + (e_{[\alpha}\rfloor
  Q_{\beta]\gamma})\,\vartheta^{\gamma} +{1\over
    2}Q_{\alpha\beta}\,.\label{N}
\end{equation}
Furthermore, it will be helpful to separate this into
\begin{equation}
N_{\alpha\beta}=N_{[\alpha\beta ]} +
\frac{1}{2}{\qslash}_{\alpha\beta}+\frac{1}{2}Qg_{\alpha\beta}\,,
\end{equation}
with $Q:=Q^\a{}_\a/n$, $\;\qslash_{\a\b}:=Q_{\a\b}-Qg_{\a\b}$, and
$g^{\alpha\beta}{\qslash}_{\alpha\beta}= 0\,$.

For $n=4$, the traceless nonmetricity
${\qslash}_{\alpha\beta}={\qslash}_{\gamma\alpha\beta}\vartheta^{\gamma}$
has $36$ independent components that can be decomposed under $O(1,3)$
as $36=16\oplus 16 \oplus 4$:
\begin{eqnarray} {\qslash}_{\alpha\beta} ={}
  ^{(1)}\!Q_{\alpha\beta}+{}^{(2)}\!Q_{\alpha\beta}
  +{}^{(3)}\!Q_{\alpha\beta}\,.
\label{Qdecomp*}
\end{eqnarray}   
Then, we have the following irreducible decomposition of the
components of the nonmetricity 1-form
$Q_{\alpha\beta}=Q_{\gamma\alpha\beta}\vartheta^{\gamma}$ with respect
to the (pseudo)-orthogonal group, cf.\ \cite{PRs,aether},
\begin{equation}\label{Qdecomp}
\underbrace{Q_{\alpha\beta}}_{\stackrel{\quad}
{\begin{picture}(25,14)(0,0)
\multiframe(1,4)(4.5,0){1}(4,4){}
\put(7,4){\tiny{$\otimes$}}
\multiframe(15,4)(4.5,0){2}(4,4){}{}
\end{picture}
}}
=\underbrace{^{(1)}Q_{\alpha\beta}}_{\stackrel{{\rm spin}\,3}
{\begin{picture}(15,10)(0,0)
\multiframe(1,4)(4.5,0){3}(4,4){}{}{}
\end{picture}
}}\,
\oplus \underbrace{^{(2)}Q_{\alpha\beta}}_{\stackrel{{\rm spin}\, 2}
{\begin{picture}(15,10)(0,0)
\multiframe(1,4)( 4.5,0){2}(4,4){}{}
\multiframe(1,-.5)(4.5,0){1}(4,4){}
\end{picture}
}}\,
\oplus \underbrace{^{(3)}Q_{\alpha\beta}}_{\stackrel{{\rm spin}\, 1}
{\begin{picture}(8,10)(0,0)
\multiframe(1,4)(4.5,0){1}(4,4){}
\end{picture}
}}\,
\oplus \underbrace{^{(4)}Q_{\alpha\beta}}_{\stackrel{{\rm spin}\, 1}
{\begin{picture}(8,10)(0,0)
\multiframe(1,4)(4.5,0){1}(4,4){}
\end{picture}
}}\quad ,
\end{equation}
where we have marked the {\em leading} spin content of the fields.  We
have also given the decomposition of the $GL(n,\mathbb R)$-reducible
components $Q_{\gamma\alpha\beta}$ into irreducible representations of
the (pseudo)-orthogonal group, so that the Young diagrams on the
right-hand-side of the above equality label $O(1,n-1)$-irreducible
representations. (Note the multiplicity 2 of the irreducible vector
representation.) The names of our corresponding computer macros are
$Q_{\alpha\beta}= {\tt trinom+binom+vecnom+conom}$. Defining 
$\Lambda_\a:= e^\b\rfloor\qslash_{\a\b}$, we have explicitly 
\begin{eqnarray}
	{}^{(1)}Q_{\a\b} &=& \Big[{\qslash}_{(\c\a\b)}-\frac{2}{n+2}\, 
  {\Lambda}_{(\c}g_{\a\b)}\Big]\vartheta^{\c}=
  \Big[{Q}_{(\c\a\b)}-\frac{2}{n+2}\,
    {\Lambda}_{(\c}g_{\a\b)}-Q_{(\c}g_{\a\b)} \Big]\vartheta^{\c}
  \,,
\label{1Q} \\
	{}^{(2)}Q_{\a\b} &=& \frac{2}{3}\,\Big[{\qslash}_{\c\a\b}-\qslash_{(\a\b)\c}
	+ \frac{1}{n-1}\,({\Lambda}_{\c}g_{\a\b}-{\Lambda}_{(\a}g_{\b)\c} )
	\Big]\vartheta^{\c}\,,
\label{2Q} \\
{}^{(3)}Q_{\a\b} &=& \frac{2n}{(n-1)(n+2)}\,\Big[{\Lambda}_{(\a}g_{\b)\c}-
\frac{1}{n}\,{\Lambda}_{\c}\,g_{\a\b}
	\Big]\vartheta^{\c}\,,
\label{3Q}  \\
{}^{(4)}Q_{\a\b} &=& g_{\a\b}Q_{\c}\vartheta^{\c}\,.
\label{4Q}
\end{eqnarray}
The irreducible part
$^{(1)}Q_{\alpha\beta}={}^{(1)}Q_{\gamma\alpha\beta}\,\vartheta^\gamma$
({\tt trinom}) corresponds to the {\it totally symmetric} piece
$^{(1)}Q_{\gamma\alpha\beta}={}^{(1)}Q_{(\gamma\alpha\beta)}$ of the
nonmetricity in which the {\it traces} have been {\it subtracted out,}
\begin{eqnarray}
  ^{(1)}Q_{\gamma\alpha\beta}
  \stackrel{n=4}{=}{\qslash}_{(\c\a\b)}-\frac{1}{3}\, 
  {\Lambda}_{(\c}g_{\a\b)}=  Q_{(\gamma\alpha\beta )} -
  \frac{1}{3}\,g_{(\gamma\alpha}\left(3Q_{\beta )}+{\Lambda}_{\beta
    )}\right)\,.
 \label{trinom***}
\end{eqnarray} 
The tracelessness of $^{(1)}Q_{\gamma\alpha\beta}$ means
$g^{\alpha\beta}\,^{(1)}Q_{\gamma\alpha\beta}=0$ and
$g^{\gamma\alpha}\,^{(1)}Q_{\gamma\alpha\beta}=0$.
The second term on the right-hand-side of
(\ref{trinom***}) takes care of
$g^{\alpha\beta}\,Q_{\gamma\alpha\beta}=nQ_\gamma$ and the third term
of $g^{\gamma\alpha}{\qslash}_{\gamma\alpha\beta}$
$=g^{\gamma\alpha}\!\!\qslash_{\gamma\beta\alpha}=\Lambda_\beta$.  The
totally symmetric piece $Q_{(\gamma\alpha\beta)}$ plays an important
r\^ole in the recent gravitational theory proposed in
\cite{Boulanger:2006tg}.  In the following we will focus on the
properties of $^{(1)}Q_{\alpha\beta}$, which, as we have seen, carries
leading spin 3.

The curvature two-form is defined by $R_{\alpha}{}^{\beta} :=
d{\Gamma}_{\alpha}{}^{\beta} - {\Gamma}_{\alpha}{}^{\g}\wedge
{\Gamma}_{\g}{}^{\beta} =\frac 12\,R_{\g\d\a}{}^\b\,\vt^\g\wedge
\vt^\d$.  Associated with it is the Ricci one-form ${\rm
  Ric}_\a:=e_\b\rfloor R_\a{}^\b={\rm Ric}_{\g\a}\,\vt^\g$. Then the
components of the Ricci tensor read $ {\rm
  Ric}_{\a\b}=R_{\g\a\b}{}^\g$. The Einstein $(n-1)$-form is given by 
$G_\a:=\frac 12\eta_{\a\b\gamma}\wedge R^{\b\gamma}$.  

With the help of (\ref{Gamma_N}) and (\ref{dist_1}) we can decompose
the total curvature $R_{\alpha}{}^{\beta}$ into Riemannian and
post-Riemannian pieces:
\begin{equation}\label{curv}
  R_{\alpha}{}^{\beta} = {\widetilde R}_{\alpha}{}^{\beta}+
  {\widetilde D}N_{\alpha}{}^{\beta} - N_{\alpha}{}^{\g}\wedge
  N_{\g}{}^{\beta}\,.
\end{equation}
In a metric-affine spacetime, the curvature 2-form can be split into a
symmetric (strain) piece $Z_{\alpha\beta}:=R_{(\alpha\beta )}$ and an
antisymmetric (rotational) piece $W_{\alpha\beta}:=R_{[\alpha\beta ]}$:
\begin{equation}
 R_{\alpha\beta}=Z_{\alpha\beta}+W_{\alpha\beta}\,.
\end{equation}
In turn, from $Z_{\alpha\beta}$, we can subtract out the trace
$Z:=Z_{\alpha}{}^{\alpha}$ and arrive thereby at the shear curvature
\begin{equation}\label{shearc} {\zslash}_{\alpha\beta}
  :=Z_{\alpha\beta}-\frac{1}{n}Zg_{\alpha\beta}\,,\quad
  {\zslash}_{\alpha}{}^\alpha=0\,.
\end{equation}

The Einstein $(n-1)$-form depends only on the rotational curvature:
\begin{equation}\label{Einstein3}
  G_\a=\frac 12\eta_{\a\b\gamma}\wedge R^{[\b\gamma]}
  =\frac 12\eta_{\a\b\gamma}\wedge W^{\b\gamma}=\frac
  12\eta_{\a\b\gamma}\wedge 
  \left({}^{(4)}W^{\b\gamma}+{}^{(5)}W^{\b\gamma}
    +{}^{(6)}W^{\b\gamma} \right)\,.
\end{equation}
If we decompose $G_\a$ respect to the $(n-1)$-form basis $\eta^\b$,
namely $G_\a=G_{\a\b}\,\eta^\b$, then the $G_{\alpha\beta}$ denote the
components of the Einstein tensor and
$G_{\alpha\beta}=W_{\g\alpha\beta}{}^{\g}-\frac{1}{2}g_{\alpha\beta}
W_{\g\d}{}^{\d\g}$.

In analogy to the Ricci one-form, we can define a Ricci-type one-form
(the ``Rizzi'' one-form) for $Z_\a{}^\b$ and $\zslash_\a{}^\b$,
respectively:
\begin{equation}\label{Rizzi}
{\rm Riz}_\a:=e_\b\rfloor Z_\a{}^\b\quad{\rm and}\quad
{\rm \Riz}_\a:=e_\b\rfloor \zslash_\a{}^\b\,.
\end{equation}
In components, we have $ {\rm Riz}_{\a\b}=Z_{\g\a\b}{}^\g$ and $
\Riz_{\a\b}=\zslash_{\g\a\b}{}^\g$.

The zeroth Bianchi identity
\begin{equation}\label{zeroBia}
DQ_{\a\b}\equiv 2Z_{\a\b}
\end{equation}
links the nonmetricity to the strain curvature. After some reordering
(see Appendix \ref{zeroB}), we can isolate a purely Riemannian covariant
derivative according to
\begin{eqnarray}\label{0thBia5b}
  &&\widetilde{D}\qslash_{\a\b}-N_{[\a\g]}\wedge\qslash^{\;\g}{}_\b
  -N_{[\b\g]}\wedge\qslash_\a{}^\g
 =2\zslash_{\a\b}\,.
\end{eqnarray}
Note that in the case of $N_{[\alpha\beta ]}=0$, the shear curvature
is completely determined by the Riemannian exterior covariant
derivative of the tracefree nonmetricity.
%
\subsection{Field equations}
%
The field equations of MAG have been derived in
a first-order Lagrangian formalism where the geometrical variables
$\{g_{\alpha\beta}\, , {\vartheta}^{\alpha}\, ,
{\Gamma}_{\alpha}{}^{\beta}\} $ are minimally coupled to matter
fields, collectively denoted ${\Psi}$, such that the total
Lagrangian, \textit{i.e.}, the geometrical part $V$ plus the matter part
$L_{{\rm matter}}$, results in
\begin{eqnarray}
	L_{{\rm total}} = V(g_{\alpha\beta}\, , {\vartheta}^{\alpha}\, ,
  Q_{\alpha\beta}\, , T^{\alpha}\, ,R_{\alpha}{}^{\beta} ) +
  L_{{\rm matter}}(g_{\alpha\beta}\, , {\vartheta}^{\alpha}\, ,
  {\Psi}\, , D{\Psi})\, .
\end{eqnarray}
Using the definitions of the excitations,
\begin{eqnarray}
M^{\alpha\beta} =
-2\frac{{\partial}V}{{\partial}Q_{\alpha\beta}}\, , \quad
H_{\alpha} = -\frac{{\partial}V}{{\partial}T^{\alpha}}\, , \quad
H^{\alpha}{}_{\beta} =
-\frac{{\partial}V}{{\partial}R_{\alpha}{}^{\beta}}\,,
\end{eqnarray}
the field equations of metric-affine gravity can be expressed
in a very concise form \cite{PRs}:
\begin{eqnarray}\label{zeroth}
  DM^{\alpha\beta} - m^{\alpha\beta} &=& \sigma^{\alpha\beta}
  \qquad\,\qquad {({\delta}/{\delta}g_{\alpha\beta})}\,,\\ DH_{\alpha}
  - E_{\alpha}\hspace{6pt} & =& \Sigma_{\alpha} \qquad\>\;\qquad {
    ({\delta}/{\delta}{\vartheta}^{\alpha})}\,,\label{first}\\
  DH^{\alpha} {}_{\beta} - E^{\alpha}{}_{\beta} &=&
  \Delta^{\alpha}{}_{\beta} \qquad\qquad{
    ({\delta}/{\delta}{\Gamma}_{\alpha}{}^{\beta})}\,,\label{second}\\
  {{\delta L}\over{\delta\Psi}} &=& 0\quad\qquad\qquad\,\; {\rm
    (matter)}\,.\label{matter}
\end{eqnarray}

As a side-remark, we discuss shortly the type of matter that couples
directly to the nonmetricity $Q_{\a\b}$, see also
\cite{Ne'eman:1996iz}.  If we go over {}from the original geometrical
variables $g_{\a\b},\vt^\a,\Gamma_\a{}^\b$ to the alternative
variables $g_{\a\b},\vt^\a,T^\a,Q_{\a\b}$, then, with the help of
Lagrangian multipliers, see \cite{PRs}, we find as response to the
variation of the torsion $T^\a$ and the nonmetricity $Q_{\a\b}$
\begin{equation}\label{varTQ}
\delta L_{\rm matter}=\cdots +\delta T^\a\wedge\mu_\a+\frac 12
\delta Q_{\a\b}\wedge\Xi^{\a\b}\,.
\end{equation}
Here the dots subsume the variations with respect to $g_{\a\b}$ and
$\vt^\a$. Hence, for the hypermomentum with its definition $\delta
L_{\rm matter}=\cdots +\delta\Gamma_\a{}^\b\wedge\Delta^\a{}_\b$, we
get
\begin{equation}\label{hyper}
\Delta^\a{}_\b=\vt^\a\wedge\mu_\b+\Xi^\a{}_\b\,,
\end{equation}
where $\tau_{\a\b}:=\Delta_{[\a\b]}$ is the spin current and the
strain-type current $ \Xi^\a{}_\b$ is symmetric: $ \Xi_{\a\b}=
\Xi_{\b\a}$.  In a hydrodynamic representation, see Obukhov and
Tresguerres \cite{hydro}, a convective ansatz for the strain-type
current reads $ \Xi_{\a\b}=\xi_{\a\b}\,\left(v\rfloor\eta \right)$,
where $v=v^\alpha e_\alpha$ is the velocity of the fluid and $\eta$
the volume $n$-form; moreover, $ \xi_{\a\b}=\xi_{\b\a}$. Accordingly,
it is the {\it material strain-type current\/} $\Xi^{\a\b}$ that couples
to the {\it nonmetricity\/} $Q_{\a\b}$. More specifically, the dilation
current $\Delta^\gamma{}_\gamma$ couples to the Weyl covector $Q$ and
the shear-type current $\Xislash:=\Xi^{\a\b}-\frac
1ng^{\a\b}\Xi^\gamma{}_\gamma$ to the tracefree nonmetricity
${\qslash}_{\alpha\beta}$.

On the right-hand-sides of each of the three
gauge field equations (\ref{zeroth}) to (\ref{second}), we identify
the material currents as sources, on the left-hand-side there are
typical Yang--Mills-like terms governing the gauge fields, their first
derivatives, and the corresponding non-linear gauge field currents.
These gauge currents turn out to be the metrical (Hilbert)
energy-momentum of the gauge fields
\begin{equation}\label{zerothx}\! m^{\alpha\beta}\! :=\!
  2{{\partial V}\over{\partial g_{\alpha\beta}}}=
  \vartheta^{(\alpha} \wedge E^{\beta )}+ Q^{(\beta}{}_{\gamma}\wedge
  M^{\alpha )\gamma}-T^{(\alpha}\wedge H^{\beta )} - R_\gamma {}^{(
    \alpha} \wedge H^{|\gamma | \beta )} + R^{( \beta |\gamma |} \wedge
  H^{\alpha )} {}_\gamma ,
\end{equation}
the canonical (Noether) energy-momentum of the gauge fields
\begin{equation}\label{firstx} E_{\alpha} :=
  {{\partial V}\over{\partial\vartheta^{\alpha}}}= e_{\alpha}\rfloor V
  + (e_{\alpha}\rfloor T^{\beta})\wedge H_{\beta} + (e_{\alpha}\rfloor
  R_{\beta}{}^{\gamma})\wedge H^{\beta}{}_{\gamma} +
  {1\over2}(e_{\alpha}\rfloor Q_{\beta\gamma})\, M^{\beta\gamma}\,,
\end{equation}
and the hypermomentum of the gauge fields
\begin{equation}\label{secondx}
  E^{\alpha}{}_{\beta}:= {{\partial
      V}\over{\partial\Gamma_{\alpha}{}^{\beta}}}= -
  \vartheta^{\alpha}\wedge H_{\beta} - g_{\beta\gamma}\,
  M^{\alpha\gamma}\,,
\end{equation}
respectively.

The {\it most general\/} parity-conserving MAG Lagrangian, at most bilinear
in $\{Q_{\alpha\beta},T^{\alpha},R^{\alpha}_{~\beta}\}$, has been
investigated by Esser \cite{Esser_dipl} and reads
\begin{eqnarray}
\label{QMA}&& V_{\rm MAG}=
\frac{1}{2\kappa}\,\Big[-a_0\,R^{\alpha\beta}\wedge\eta_{\alpha\beta}
  -2\lambda_{0}\,\eta
  \cr &&\hspace{20pt}
  +\;T^\alpha\wedge{}^\star \Big(\sum_{I=1}^{3}a_{I}\,^{(I)}
    T_\alpha\Big) + Q_{\alpha\beta} \wedge{}^\star
  \Big(\sum_{I=1}^{4}b_{I}\,^{(I)}Q^{\alpha\beta}\Big)
\nonumber \\ &&\hspace{20pt}
+\;
  2\Big(\sum_{I=2}^{4}c_{I}\,^{(I)}Q_{\alpha\beta}\Big)
  \wedge\vartheta^\alpha\wedge{}^\star \!\, T^\beta + b_{5}
  \left(^{(3)}Q_{\alpha\gamma}\wedge\vartheta^\alpha\right)\wedge
  {}^\star \!\left(^{(4)}Q^{\beta\gamma}\wedge\vartheta_\beta
  \right)\Big]\\ 
  & &\hspace{20pt} -\frac{1}{2\rho}\,R^{\alpha\beta}
\wedge{}^\star \!
\left[\sum_{I=1}^{6}w_{I}\,^{(I)}W_{\alpha\beta} +
  \sum_{I=1}^{5}{z}_{I}\,^{(I)}Z_{\alpha\beta} \nonumber\right.\\
&&\hspace{20pt} \left.
+\;w_7\,\vartheta_\alpha\wedge(e_\gamma\rfloor
  ^{(5)}W^\gamma{}_{\beta} ) +z_6\,\vartheta_\gamma\wedge
  (e_\alpha\rfloor ^{(2)}Z^\gamma{}_{\beta}
  )+\sum_{I=7}^{9}z_I\,\vartheta_\alpha\wedge(e_\gamma\rfloor
  ^{(I-4)}Z^\gamma{}_{\beta} )\right]\,.\nonumber
\end{eqnarray}
One should also consult
Refs.\cite{Esser_dipl,Yuri_effective,MAGII,aether} and the literature
quoted there.

Here $\kappa$ is the dimensionful ``weak'' Newton--Einstein
gravitational constant, $\lambda_{0}$ the ``bare'' cosmological
constant, and $\rho$ the dimensionless ``strong'' gravity coupling
constant. The constants $ a_0, \ldots a_3$, $b_1, \ldots b_5$, $c_2,
c_3,c_4$, $w_1, \ldots w_7$, $z_1, \ldots z_9$ are dimensionless and
give a weight for the different contributions of
each linearly-independent term entering the Lagrangian.\\
Actually we will not consider the complete Lagrangian (\ref{QMA}).
Instead, we choose a simplified version with
\begin{equation}\label{exotic}
  w_7=z_6=z_7 =z_8 =z_9=0 
\end{equation}
whose effect is to decouple $Z^{\alpha}_{~\beta}$ {}from $W^{\alpha}_{~\beta}$
in the Lagrangian. 
Taking (\ref{exotic}) into account, the various excitations
$\{M^{\alpha\beta}\,,H_{\alpha}\,, H^{\alpha}{}_{\beta}\}$ are
found to be
\begin{eqnarray}\label{M-excit}
  M^{\alpha\beta} & = & -\frac{2}{\kappa} ^{\star }\! \left(
    \sum\limits_{I=1}^{4} b_{I} {^{(I)}Q}^{\alpha\b} \right) \cr & & \cr
  & & -\frac{2}{\kappa}\left[ c_{2}{\vartheta}^{({\alpha}}\wedge {}
    ^{\star (1)}T^{\beta )} + c_{3}{\vartheta}^{({\alpha}}\wedge {}
    ^{\star (2)}T^{\beta )} + \frac{1}{4}(c_{3}-c_{4})\,^{\star
      }Tg^{\alpha\beta} \right] \cr & & \cr & &
  -\frac{b_{5}}{\kappa}\left[ {\vartheta}^{( \alpha}\wedge
    {}^{\star}(Q\wedge {\vartheta}^{\beta )})-\frac{1}{4}
    g^{\alpha\beta}{}\,^{\star}(3Q+{\Lambda})\right]\,,\\&&\cr
  H_{\alpha}& =& -\frac{1}{\kappa} ^{\star }\!\left(
    \sum\limits_{I=1}^{3} a_{I} {^{(I)}T}_{\alpha} +
    \sum\limits_{K=2}^{4} c_{K} {^{(K)}Q}_{\alpha\beta}\wedge
    {\vartheta}^{\beta}\right)\,,
\label{Ha-excit}\\ &&\cr\label{Hab-excit}
H^{\alpha}{}_{\beta}&=&\frac{a_{0}}{2\kappa}{\eta}^{\alpha}{}_{\beta}+
\sum\limits_{I=1}^{6}w_{I}{}^{\star (I)}W^{\alpha}{}_{\beta} +
\sum\limits_{K=1}^{5}z_{K}{}^{\star (K)}Z^{\alpha}{}_{\beta}\, .
\end{eqnarray}
The general structure of the excitations can be found in \cite{PRs},
compare also \cite{Esser_dipl}.

\section{Massless spin-3 theory in MAG }
\label{Vasiliev}
%
In this section, we show that, as was expected {}from the decomposition 
(\ref{Qdecomp}) of $Q_{\alpha\beta}$, the action of MAG in the free 
limit and in Minkowski spacetime indeed incorporates Fronsdal's action for a 
massless spin-3 gauge field, the latter field being dynamically represented by
$^{(1)}Q_{\alpha\beta}$. 
\vspace*{.1cm}

As was first shown by Fronsdal \cite{Fronsdal:1978rb} in 1978, a
massless integer-spin gauge field in Minkowski spacetime is described
by a totally symmetric tensor $h_{ i_1\ldots i_s}$ subject to the
double tracelessness condition (for $s\geqslant 4$) $o^{ i_1 i_2}o^{
  i_3 i_4}h_{ i_1\ldots i_s}=0\,$.  A quadratic Lagrangian for a free
spin-$s$ field is fixed unambiguously in the form $L^s = h \vec{L}
h\,$ (where $\vec{L}$ is some second-order differential operator) by
the requirement of gauge invariance under the Abelian gauge
transformations $\delta h_{ i_1\ldots i_s}=s\;\pa_{( i_1}\l_{
  i_2\ldots i_s)}$.  The gauge parameters $\l_{ i_1\ldots i_{s-1}}$
are rank-$(s-1)$ totally symmetric traceless tensors, with $\l_{
  i_1\ldots i_{s-1}}=\l_{( i_1\ldots i_{s-1})}$ and $o^{ i_1 i_2}\l_{
  i_1\ldots i_{s-1}}= 0\,$.  This formulation is parallel
\cite{deWit:1980pe} to the metric formulation of gravity.

In 1980, in view of extending supergravity theories by the addition of 
high-spin gauge fields, Vasiliev proposed a frame-like reformulation  
of Fronsdal's theory by using generalized vielbeins and spin connections 
\cite{Vasiliev:1980as}. 

In the next subsections, we briefly review Fronsdal's and Vasiliev's
approaches for the massless spin-3 gauge field.  Both approaches will
be needed when showing the occurrence of a massless spin-3 sector in
MAG.

\subsection{Massless spin-3 field in Fronsdal's approach}
\label{sec:Frons3}

The action given in \cite{Fronsdal:1978rb} for a totally symmetric
massless spin-$3$ gauge field $h_{ijk}=h_{(ijk)}$ in Minkowski
spacetime reads
\begin{eqnarray}
S[h_{ijk}]= - \frac{1}{2}\, \int d^n x\hspace{-7pt} &\Big[\hspace{-7pt}&
\partial_{\ell}h_{ijk }\,\partial^{\ell}h^{ijk }
- 3 \;\partial_{j}h^{\ell}_{~\,\ell i}
\,\partial^{j}h_{k }^{~\,k  i}
+6\; \pa_{j}h^{\ell}_{~\,\ell i}\,\partial_{k}h^{j k i}
\nonumber \\
&&-\;3\;\partial_{j}h^{j}_{~\,ik }\,\pa_{\ell}h^{\ell ik }
-\frac{3}{2}\;\partial^{j}h^{\ell}_{~\,\ell j}\,
\pa_{i}h^{k \,~i}_{~\,k } \Big]\,.
\label{FronsA}
\end{eqnarray}
It is invariant under the gauge transformations 
\begin{eqnarray}
	\delta h_{ijk } = 3\,\partial_{(i}\lambda_{jk )}\,,
	\quad\lambda_{ij}=\lambda_{(ij)}\,,\quad
  o^{ij}\lambda_{ij}  =  0\,.
\label{FronsGT}
\end{eqnarray}
The corresponding source-free field equations are equivalent to 
\begin{eqnarray}
{\cal F}_{ijk } := \Box h_{ijk } -
3\;\partial^{\ell}\partial_{(i}h_{jk )\ell} +
3\;\partial_{(i}\partial_{j}h_{k )\ell}^{~~~\ell} = 0\,.
\label{frspin3}	
\end{eqnarray}
It is possible to reach the harmonic gauge 
\begin{eqnarray}
	D_{jk } := \partial^{i}h_{i jk } 
	-\partial_{(j}h_{k)i}^{~~~i} = 0\,,\qquad
	\delta D_{jk } = \Box \lambda_{jk }
\end{eqnarray}
in which the field equations take the canonical massless Klein--Gordon
form $\Box h_{ijk }=0\,$.  By a residual gauge transformation with
parameter $\bar\lambda_{ij}$ obeying $\Box \bar\lambda_{ij}=0$, it is possible
to set the trace of the gauge field to zero, yielding
\begin{eqnarray}
  \Box h_{ijk} = 0\,,\quad \partial^{i}h_{ijk}=0\,, \quad o^{ij}h_{ijk} = 0 \, .
	\label{dWF}
\end{eqnarray}
Actually, some residual gauge transformations $\delta
h_{ijk}=3\,\partial_{(i}\tilde{\lambda}_{jk)}$ are still allowed in
(\ref{dWF}).  As shown in \cite{deWit:1980pe}, this gauge theory leads
to the correct number of physical degrees of freedom, that is, to the
dimension of the irreducible representation of the little group
$O(n-2)$ corresponding to the one-row Young diagram of length $s=3\,$.

The counting of physical degrees of freedom can also be done by using
the gauge-invariant spin-3 Weinberg tensor ${\cal K}$
\cite{Weinberg:1965rz} (see also \cite{deWit:1980pe}) which is the
projection of $\partial_{i}\partial_{k}\partial_{m}h_{n \ell j}$ on
the tensor field irreducible under $GL(n,\mathbb R)$ with symmetries
labeled by the Young tableau
\begin{picture}(43,16)(0,0) \multiframe(1,4)(12.5,0){3}(12,12){\small
    $i$}{\small $k$}{\small $m$}
  \multiframe(1,-8.5)(12.5,0){3}(12,12){\small $j$}{\small $\ell$}{\small $n$}
\end{picture}\,.
Since $\partial_{i}\partial_{k}\partial_{m}h_{n \ell j}$ is
already symmetric in all indices of the two rows of the above Young
tableau, it only remains to antisymmetrize over
the three pairs $(ij\,,k\ell\,,mn)$.
This corresponds to taking $3$ curls of the symmetric tensor field
$h_{n\ell j}\,$ and yields a curvature-like tensor
\begin{equation}\label{defK} {\cal
    K}_{ij\,k\ell\,mn}:=8\partial_{[i|}\partial_{[k|}\partial_{[m}h_{n]|\ell
    ]|j]}\,.
\end{equation}
In fact, the source-free Fronsdal equations (\ref{frspin3}) imply the
Ricci-flat-like equations
\begin{eqnarray}
	{\cal F}=0\quad \Rightarrow\quad {\rm Tr}\, {\cal K} = 0\quad
        \Leftrightarrow\quad o^{ik}{\cal K}_{ij\,k\ell\, mn} = 0\,.
\label{Ricciflat}
\end{eqnarray} 
Conversely, it was shown in \cite{Bekaert:2003az} that the
Ricci-flat-like equations Tr ${\cal K}=0$ imply\footnote{ More
  details, references and general results for tensor gauge fields
  transforming in arbitrary irreducible representations of $GL(n,{
    \mathbb R } )$ can be found in \cite{Bekaert:2006ix}.  Note that
  by introducing a pure gauge field (sometimes referred to as
  ``compensator"), it is possible to write a local (but
  higher-derivative) action for spin-$3$ \cite{FS} that is invariant
  under unconstrained gauge transformations.  Recently, this action
  was generalized to the arbitrary spin-$s$ case by further adding an
  auxiliary field \cite{Francia:2005bu} (see also
  \cite{Pashnev:1998ti} for an older non ``minimal" version of it).}
the Fronsdal equations ${\cal F}=0\,$.  This was obtained by combining
various former results \cite{Damour:1987vm,DVH,FS}.  Using the
definition of ${\cal K}$, the equations (\ref{Ricciflat}) give the
following set of first-order field equations:
\begin{eqnarray}
\left\{
\begin{array}{c}
\partial_{[i}{\cal K}_{j k]\,\ell m \, no}=0\,, \\
\;\;\partial^i {\cal K}_{i j\, k \ell \,mn} = 0\,,
    \end{array}\right.\qquad \quad {\rm where~~Tr\,} {\cal K} = 0\,.
    \label{BWequs}
\end{eqnarray}
When $n=4$, the above equations correspond to the (spin-3) 
Bargmann--Wigner equations~\cite{Bargmann:1948}, originally
expressed in terms of two-component tensor-spinors in the
representation $(3,0)\oplus(0,3)$ of $SL(2,\mathbb C)$. 
See also \cite{Damour:1987vm} for a careful analysis
of Fronsdal's spin-3 gauge theory using the Weinberg tensor ${\cal K}$
(denoted $R_6$ in \cite{Damour:1987vm}). 

In the massless spin-1 case, the Bargmann--Wigner equations read
\begin{eqnarray}\label{Max}
\left\{
\begin{array}{c}
 \partial_{[i}F_{jk]}=0\,,
	\\
\;\;\partial^{i}F_{ij} = 0\,,
    \end{array}\right.
\end{eqnarray}
which are nothing but the source-free Maxwell equations. They imply 
$\Box F = 0$ and $F = dA$, where as usual $F=\frac{1}{2}\,F_{ij}\,dx^i\wedge dx^j$
and $A=A_idx^i\,$. They are invariant under $\delta A = d\lambda\,$. 
The tensor $F$ transforms in the representation $(1,0)\oplus (0,1)$ of 
$SL(2,\mathbb C)$. 
One can choose the Lorentz gauge-fixing condition $\partial^i A_i = 0$ 
and look for solutions of the source-free Maxwell equations with the ansatz
\begin{eqnarray}
  A=A_{i}dx^i = \Phi(x)k=\Phi(x)k_idx^i\,,
\label{an1}
\end{eqnarray}
where $k_{i}$ are the constant components of a 1-form $k$, which is
null: $k\wedge{}^\star k=0$.  We may choose the vector dual to the
1-form $k$ in the $z$-direction: $k^{i}=(E,0,0,E)$.  The Lorentz
condition ${}^{\star}d\,{}^{\star}A=0$ implies the equation
$k\wedge{}^{\star}d\Phi = 0$, which is satisfied with
$\Phi=\phi(\xi^a)\,e^{ik\cdot x}$ where $k\cdot x :=
k_{i}x^{i}=-Et+Ez$ and where $\phi(\xi^a)$ is a function of the
transverse coordinates $\xi^1=x$, $\xi^2=y\,$. [Implicitly, the real
component of $\Phi$ must be taken.]  Then, the d'Alembert equation
$(d\,{}^{\star} d{}^{\star}+{}^{\star} d{}^{\star}d )A = 0$ is
verified if $\phi(\xi^a)$ is a harmonic function in the $(x,y)$-plane,
$\Delta_{x,y}\phi(\xi^a)=0$.  The monochromatic plane-wave solution
$A_{i}=\phi(\xi^a)\,k_{i}\,e^{i k\cdot x}$ displayed here
characterizes a pure-radiation electromagnetic field $F$ (also called
{\emph{null field}}) since we have the vanishing of the two invariants
$F\wedge F$ and $F\wedge{}^{\star}F\,$.  Note also that we have
$A\wedge dA = 0$, which implies by Frobenius's theorem that the vector
dual to $A$ is hypersurface orthogonal, the surface being described by
the equation $\Sigma\equiv k\cdot x +\text{const} = 0\,$.
 
With the pure-radiation massless spin-1 solution $F=dA$ displayed above, 
it is simple to construct helicity-$3$ plane-wave solutions of
the Bargmann--Wigner equations (\ref{BWequs}): 
\begin{eqnarray}
	h_{ijk} = \Phi\,k_{i}k_{j}k_{k}\,,
	\quad k_{i}=(-E,0,0,E)\,,\quad \Phi=\phi(\xi^a)\,e^{ik\cdot x}\,,\quad 
	\Delta_{\xi}\phi(\xi^a)=0=\Box \Phi\,.
	\label{solus} 
\end{eqnarray}
Indeed, computing the spin-3 Weinberg tensor ${\cal K}_{ij\,k\ell\,mn}$, 
we find 
\begin{equation}\label{kkk}{\cal K}_{ij\,k\ell\,mn} = 
-8\left(k_{[i|}k_{[k|}k_{[m}\partial_{n]}\partial_{|\ell]}
\partial_{|j]}\phi\right)\,e^{ik\cdot x}\,.
\end{equation}
By using the properties of $k$ and $\phi\,$, it can be shown that the
Bargmann--Wigner equations (\ref{BWequs}) are obeyed.  Hence, on-shell,
the field strength ${\cal K}$ is a propagating massless helicity-$3$
field. It gives a representation of $SL(2,\mathbb{C})$ labeled by
$(0,3)\oplus (3,0)$ and satisfies the massless Klein--Gordon equation
$\Box {\cal K}=0\,$.  In the van der Waerden 2-spinor notation, the
monochromatic plane-wave solution written above corresponds to a $\cal
K$ that is equivalent to a totally symmetric $6$-spinor with all $6$ null
directions coinciding. The (6 times repeated) null spinor represents
the light-like wave covector $k_{i}\,$, cf.\ \cite{Damour:1987vm}.
Finally note that (i) the equations (\ref{BWequs}) hold in arbitrary
dimension $n>2\,$ and (ii) the gauge potentials $h_{ijk}$ given
in (\ref{solus}) satisfy the equations (\ref{dWF}).

Actually, we can put the plane-wave solutions (\ref{solus}) in exactly
the same form as that found by Obukhov \cite{Yuripp-waves2006} for
metric-affine gravity, see also Pasic and Vassiliev
\cite{Pasic:2005qr}. One must identify Obukhov's 1-form $u$ with our
1-form $k\Phi$ and his $H$ with our $\Phi$, so that the nonmetricity
reads $Q_{\a\b}=k_{\a}k_{\b}u=\Phi\, k_{\a}k_{\b}k\,$.  Then, as done
in \cite{Yuripp-waves2006}, it is straightforward to add torsion by
taking $\Gamma_{\alpha\beta}=k_{[\a}\varphi_{\b]}k+k_{\a}k_{\b}u$,
where $\varphi_{\a}=\partial_{\a}H\,$.  Similarly, one can choose the
coframe and metric as in \cite{Yuripp-waves2006}, since they only
depend on the function $H\,$.  The only component of the curvature
$W_{\a\b}$ that remains is the Weyl piece $^{(1)}\!W_{\a\b}\,$.  In
conclusion, with the identifications explained here, we have made the
exact correspondence between our plane-wave solutions (\ref{solus})
and those of Obukhov \cite{Yuripp-waves2006}.

\subsection{ Vasiliev's approach to massless spin-3 field}

Fronsdal's action for a massless spin-$s$ gauge field in Minkowski
spacetime was elegantly rewritten by Vasiliev \cite{Vasiliev:1980as}
in a first-order frame-like formalism.  In the particular spin-3 case,
the set of bosonic fields consists of a generalized vielbein $e_{i\,
  \alpha\beta}$ and a generalized spin connection $\o_{i\,\g \a\b}\,$.
They obey the following algebraic identities
\begin{eqnarray}
e_{ i\, \alpha\beta}& =& e_{ i\,\beta\alpha}\,,\hspace{45pt}
 o^{\alpha\beta}e_{ i\,\alpha\beta} =0\,,
\nonumber \\
\o_{ i\, \g\, \a\b}&=&\o_{ i\, \g\, \b\a}
\,,\qquad \o_{ i\, (\g\, \a\b)}=0\,,\label{algcond1}
 \\
 o^{\alpha\b}\o_{ i\, \g \a\b}&=&0\,,\hspace{40pt}
o^{\g\alpha}\o_{ i\, \g\alpha \b}=0\,. 
 \nonumber	
\end{eqnarray}
The action was originally written in four dimensions as \cite{Vasiliev:1980as}
\begin{eqnarray}
	S[e,\o]=\int d^4x \, \varepsilon^{ijk\ell} \,
\varepsilon_{\alpha \b\g k} 
\, \omega_\ell{}^{\a\b \d}\Big(
\pa_{i}e_{j\, \d}{}^{\g}
-\frac12\,\omega_{i\,j \d}{}^{\g}\Big).
\label{Vasaction}
\end{eqnarray}
As in the Einstein--Cartan theory of gravitation (see \cite{Erice95}
and \cite{Trautman}), the connection is a non-propagating field. One
can solve the source-free field equations for $\omega_{i\,\g\a\b}$ and
express it in terms of the frame-like field $e_{i\,\alpha\beta}$.
Inserting the result back in the action (\ref{Vasaction}) and
multiplying by $1/\rho$ for further purpose, one obtains an action in
second-order formalism, in a form valid in any number of spacetime
dimensions\footnote{See Eq.\ (20) of \cite{Aragone:1981yn} with the
  identification $f_{mn\tilde{a}\tilde{b}} \rightarrow
  \czslash_{mn\a\b}\,$; see also \cite{Boulanger:2003vs} with
  $B_{mn|ab} \rightarrow \czslash_{mn\a\b}$.}
\begin{eqnarray}\label{Va}
  S[{e}_{\g\a\b}] &=& \frac{1}{\rho}\,
  \int d^nx \Big[ (\frac{1}{4}\czslash_{\g\d\a\b} 
  -\czslash_{\a\g\d\b} )
  \czslash^{\,\g\d\a\b} -( 2\czslash_{\g\a\b}{}^\g+
  \czslash_{\g\b\a}{}^\g) \czslash_\d{}^{\a\b\d}{}\Big]\,,
\end{eqnarray}
where $\czslash_{\alpha\beta}=\frac{1}{2}
\czslash_{\g\d\alpha\beta}\,\vt^\g\wedge\vt^{\d}=\frac{1}{2}
\czslash_{ij\alpha\beta}\,dx^i\wedge dx^j$ is the curvature-like
two-form constructed {}from the one-form
${e}_{\alpha\beta}={e}_{i\alpha\beta}\,dx^i={e}_{\g\alpha\beta}\,\vt^\g$
by exterior differentiation:
\begin{eqnarray}
\czslash_{\alpha\beta}:=d {e}_{\alpha\beta} \qquad \Leftrightarrow \qquad
\czslash_{\g\d\alpha\beta} = 2 \,\pa_{[\g}{e}_{\d]\alpha\beta}\,.
\label{czsl}
\end{eqnarray}
Note that $\partial_\g:=e^i{}_\g\partial_i$. The action (\ref{Va}) is
invariant under the gauge transformations
\begin{eqnarray}\label{Vasymm}
    \d {e}_{\g \alpha\beta} = \pa_{\g} \hat{\xi}_{\alpha\beta}
  + \hat{a}_{\g  \alpha\beta}\,,
\end{eqnarray}
where $\hat{a}_{\g\alpha\beta}$ is traceless,
$o^{\a\b}\hat{a}_{\g\alpha\beta}=0$ and
$o^{\g\alpha}\hat{a}_{\g\alpha\beta}=0$, and transforms in the
$(2,1)$-module of ${O}(1,n-1)$ denoted by the Young tableau
{\scriptsize{
\begin{picture}(25,15)(-5,2)
\multiframe(0,7.5)(7.5,0){1}(7,7){$\alpha$}
\multiframe(7.5,7.5)(7.5,0){1}(7,7){$\beta$}
\multiframe(0,0)(7.5,0){1}(7,7){$\g$}
\end{picture}}}.
The gauge parameter
$\hat{\xi}_{\alpha\beta}$ is symmetric $\hat{\xi}_{\alpha\beta}=
\hat{\xi}_{\b\a}$ and traceless
$o^{\alpha\beta}\hat\xi_{\alpha\beta}=0$, \ie it transforms in the
$(2,0)$-module\hspace{-5pt} {\scriptsize{
    \begin{picture}(25,15)(-5,7)
      \multiframe(0,7.5)(7.5,0){1}(7,7){$\alpha$}
      \multiframe(7.5,7.5)(7.5,0){1}(7,7){$\beta$}
\end{picture}
}}\hspace{-5pt} of ${O}(1,n-1)$.

Due to the gauge symmetry $\d_{\hat{a}} {e}_{\g \alpha\beta} =
\hat{a}_{\g \alpha\beta}$, only the totally symmetric component of
${e}_{\g\alpha\beta}$ survives in the action, yielding Fronsdal's
action (up to an inessential overall constant factor) for
$h_{\g\alpha\beta}\equiv e_{(\g\alpha\beta)}\,$, invariant under
$\delta {h}_{\g \alpha\beta} =
3\,\partial_{(\g}\lambda_{\alpha\beta)}$, with
$\lambda_{\alpha\beta}=\frac{1}{3}\,\hat{\xi}_{\alpha\beta}\,
$~\cite{Vasiliev:1980as}.
%
\subsection{Fronsdal's action in MAG}
%
As we anticipated by using the notation $\czslash_{\alpha\beta}$ for
the curvature-like two-form of Vasiliev's spin-3 vierbein one-form
$e_{\alpha\beta}$, the Lagrangian in (\ref{Va}) is contained in a
general MAG Lagrangian (\ref{QMA}) taken at quadratic order and
evaluated in flat spacetime.  The crucial point is to identify
Vasiliev's spin-3 frame field with the traceless nonmetricity:
\begin{eqnarray}
  e_{\alpha\beta} = \frac{1}{2}\qslash_{\alpha\beta}& =& \nonumber
  \Gamma_{(\alpha\beta)}-\frac{1}{n}\,g_{\alpha\beta}\,\Gamma_{\g}{}^\g-
  (\widetilde{\Gamma}_{(\alpha\beta)}-\frac{1}{n}\,g_{\alpha\beta}\,
\widetilde{\Gamma}_{\g}{}^\g)\\ &\stackrel{*}{=}&
 \Gamma_{(\alpha\beta)}-\frac{1}{n}\,o_{\alpha\beta}\,\Gamma_{\g}{}^\g  \,,
\label{identif}	
\end{eqnarray}
where the tilde denotes the Riemannian connection and the star refers
to orthonormal coordinates.  Then, taking the traceless part of the
zeroth Bianchi identity $\frac{1}{2}\,DQ_{\alpha\beta}\equiv
Z_{\alpha\beta}$ and recalling the definition (\ref{shearc}) of the
shear curvature $\zslash_{\a\b}$, one finds (the irreducible
decomposition is listed in Appendix \ref{irrdec.Z}):
\begin{eqnarray}
  \frac12\, D \qslash_{\a\b}\equiv\zslash_{\a\b} ={}  ^{(1)}\zslash_{\a\b}+{}^{(2)}
  \zslash_{\a\b}
  +{}^{(3)}\zslash_{\a\b}+{}^{(5)}\zslash_{\a\b}\,.
\end{eqnarray}
This is an exact relation valid in each metric-affine space.
If we now use {\it orthonormal} coordinates and {\it linearize}, we
discover that
\begin{eqnarray}
	\frac12\, d \qslash_{\a\b}=de_{\a\b}={}  ^{(1)}\czslash_{\a\b}+{}^{(2)}
  \czslash_{\a\b}
  +{}^{(3)}\czslash_{\a\b}+{}^{(5)}\czslash_{\a\b}\,.
\end{eqnarray}
Here $\czslash_{\a\b}$ is the curvature-like two-form defined in
(\ref{czsl}).  Of course, since the decomposition of
$\zslash_{\alpha\beta}$ is purely algebraic, it also holds at the
linearized level, for $\czslash_{\alpha\beta}$.

We can now equate the Lagrangian (\ref{QMA}) with (\ref{Va}) (the
former taken at quadratic order, in Minkowski spacetime). We obtain a
system of linear equations for the parameters $a_0$, $\ldots$, $a_3$,
$b_1$, $\ldots$, $b_5$, $c_2$, $c_3$, $c_4$, $w_1$ $\ldots$, $w_7$,
$z_1$ $\ldots$, $z_9$.  Obviously, only the terms $\int
DZ\wedge{}^{\star} DZ$ of (\ref{QMA}) will contribute to the action
(\ref{Va}), so that only the constants $z_1,z_2,z_3,z_4,z_5$ will be
nonzero \textit{a priori}.  Furthermore, one can already guess that
$z_4$ will be vanishing because Vasiliev's action (\ref{Va}) involves
only the traceless part $\czslash_{\alpha\beta}$ of $\cz_{\a\b}$,
which is linearly independent {}from the pure trace part
${}^{(4)}\cz_{\alpha\beta}$.

Using Appendix \ref{irrdec.Z}, the volume $n$-form $\eta$, and the
Rizzi-like one-form associated with $\czslash_{\a\b}$ [namely
$\cRiz_{\a}$, cf.\ (\ref{Rizzi})], the Lagrangian $\cl
=L\eta=-\frac{1}{2\rho}\,\czslash^{\alpha\beta} \wedge{}^\star \Big(
\sum\limits_{I=1,2,3,5} {z}_{I}\,^{(I)}\czslash_{\alpha\beta} \Big)$
can be written as
\begin{eqnarray}
  L &=&  \frac{z_1
    +z_2}{8\rho}\czslash_{\g\d\alpha\beta}\czslash^{\,\g\d\alpha\beta}
  - \frac{z_1-z_2}{4\rho}\czslash_{\g\d\alpha\beta}\czslash^{\,\alpha\g\d\beta}
  \nonumber \\
  &&\;- \frac{1}{4{\rho}}\left[
    \frac{3n+4}{n(n+2)}z_{1}+\frac{z_{2}}{n-2}-\frac{2n}{n^2-4}z_{3}
    -\frac{2}{n}z_{5}\right]
  \cRiz_{\alpha\beta}
  \cRiz^{\a\beta}
  \nonumber \\
  &&\;-\frac{1}{4{\rho}}\left[ \frac{n+4}{n(n+2)}z_{1} -
    \frac{z_{2}}{n-2}+ \frac{2n}{n^2-4}z_{3}-\frac{2}{n}z_{5}\right]
  \cRiz_{\alpha\beta}
  \cRiz^{\b\a}\,.
           \label{Ldecomp}
\end{eqnarray}
Hence (\ref{Ldecomp}) is equal to the Lagrangian in (\ref{Va}) 
if and only if the following equations hold:
\begin{eqnarray}
\label{valuesVa}
  z_1 = 3\,,\quad z_2 = -1\,,\quad z_3 = 1-n\,,\quad z_5 = 3(1-n)\,,
\end{eqnarray}
all the other constants, in particular $z_4$, being equal to zero.
Accordingly, Vasiliev's action (\ref{Va}) reads
\begin{eqnarray}
S^{\rm Vasiliev}[e_{\g\a\b}]=S^{\rm Fronsdal}[h_{\g\a\b}]=
-\frac{1}{2\rho}\int_{\Omega_n} \,\czslash^{\alpha\beta}
\wedge{}^\star
\Big(
  \sum_{I=1,2,3,5} {z}_{I}\,^{(I)}\czslash_{\alpha\beta} \Big)	
\end{eqnarray}
together with (\ref{valuesVa}).  
Finally, the field equations turn out to be
\begin{eqnarray}\label{fieldeq}
0=\frac{\delta S^{\rm Vasiliev}}{\delta e^{\g\a\b}}=
-\frac{1}{\rho}\;\partial^{\delta}\!\!&\Big[&
2\,\czslash_{\alpha[\g\delta]\beta}+2\,\czslash_{\beta[\g\delta]\alpha}
- \czslash_{\g\delta\alpha\beta}
+4\, o_{\beta[\g}\cRiz_{\delta]\alpha}
+4\, o_{\alpha[\g}\cRiz_{\delta]\beta}
\nonumber \\
&{}&+\;\,
2\,o_{\g(\alpha}\cRiz_{\beta)\delta}-
2\,o_{\delta(\alpha}\cRiz_{\beta)\g}
\Big]\,.
\end{eqnarray}
Because of the equality
$S^{\rm{Fronsdal}}=S^{\rm{Vasiliev}}$, the equations (\ref{fieldeq})
are equivalent to Fronsdal's equations (\ref{frspin3}).

It is possible to pick up a gauge in which the only irreducible part
that remains of the shear curvature $\czslash_{\g\d\alpha\beta}$ is
its first component ${}^{(1)}\czslash_{\g\d\alpha\beta}\propto
\partial_{[\g}h_{\d]\alpha\beta}$.  The field $h_{\a\b\g}\equiv
e_{(\a\b\g)}$ is the only component of the frame-like field that
survives in the action, while the trace $o^{\a\b}h_{\a\b\g}$ and
the divergence $\partial^{\a}h_{\a\b\g}$ both vanish in the
appropriately chosen gauge.  This gauge is the one for which the field
equations take the form (\ref{dWF}).  As noted at the end of Section
\ref{sec:Frons3}, the plane-wave solutions (\ref{solus}) satisfy the
corresponding gauge conditions.  Therefore, it is easy to see that the
components ${}^{(I)}\!\!\czslash_{\g\d\alpha\beta}\,$, $I=2,3,5\,$,
are zero for the plane-wave solutions constructed in
(\ref{solus}).

Up to an inessential factor of $2$, we have thus identified the spin-3
gauge field in Vasiliev's frame formalism with the component of the
nonmetricity one-form which lies along the shear generator of
$GL(n,\mathbb{R})$.  This enabled us to show in a direct way the
appearance of Fronsdal's massless spin-3 action as a part of MAG's
action (\ref{QMA}), provided that the free parameters present in the
latter action are picked according to (\ref{valuesVa}), the remaining ones
being zero altogether.
%
%
\section{Spin-3-like exact solutions of full nonlinear MAG}
\label{sec:exsol}
%
As we have shown in the previous section, in the gravitational gauge
sector of MAG, the connection ${\Gamma}_{\alpha}{}^{\beta}$ already
mediates particles of different spin content, {}from 1 to 3.  Since
the works of Fronsdal \cite{Fronsdal:1978rb,FangF,Fronsdal:1978vb}, it
has been widely recognized that free massive and massless higher-spin
fields consistently propagate in maximally symmetric
spaces\footnote{See the recent work of Buchbinder et al.\
  \cite{Buchbinder:2006ge} for more details and references.  See also
  Illge \& Schimming \cite{Illge:1999tb}, Illge \& W\"unsch
  \cite{IllgeWuensch}, and references therein, where more general
  backgrounds have been investigated.}, and consistent higher-spin
cubic vertices have been obtained in such spaces (see
\cite{Metsaev:2005ar} for a light-cone analysis and references on the
problem of consistent higher-spin cubic vertices, including
Yang--Mills and gravitational couplings; see
\cite{Bekaert:2005jf,Boulanger:2005br} for non-Abelian massless spin-3 
covariant cubic vertices in flat space; higher-derivative Abelian vertices 
are discussed in \cite{Damour:1987fp}).  
However, so far, no interacting
Lagrangian --- consistent at all orders in the coupling constants ---
has yet been written that would non-trivially involve spin-3 gauge
fields. Presumably an infinite number of higher-spin fields is
required.  The best hope in that direction is the theory initiated by
Fradkin and Vasiliev \cite{FV}, further developed notably in \cite{V}
and reviewed, e.g., in \cite{Bekaert:2005vh}.
  
In the field theoretical approach proposed in
\cite{Kirsch:2005st,Boulanger:2006tg}, higher-spin connections arise
in the context of symmetry breaking mechanisms starting {}from the
group of analytical diffeomorphism $G=Di\!f\!f(n,R)$. Breaking this
symmetry down to the Lorentz group $SO(1,n-1)$, e.g., those
generalized connections can be identified with certain parameters of
the coset space $G/H$ and give rise to an infinite tower of 
higher-spin fields, cf.\ also~\cite{PRs,Lopez_Tresguerres_1995,Tiemblo:2005js}.
 
Because of the identification (\ref{identif}) and the results of the
previous section, it appears that full nonlinear MAG offers an
interesting vantage point on the difficult problem of spin-3
interactions, with itself and with gravity. Therefore, an important
step in that direction is to search for exact solutions of full
nonlinear MAG that propagate the spin-3 field $^{(1)}Q_{\alpha\beta}$.
Moreover, such exact solution are, within MAG, interesting for their
own sake, in particular also for studying non-Riemannian cosmological
models, see Puetzfeld \cite{Puetzfeld2004a,Puetzfeld2004c}.

\subsection{Ansatz for the nonmetricity}
\label{ANSATZ}
%
To isolate the main spin-3 content of the connection, we will
postulate the existence of a 1-form ${\bf \ell}(x)$ and a scalar field
${\Phi}(x)$, such that the nonmetricity can be parameterized according to
\begin{equation}\label{ansatz_Q1}
Q_{\alpha\beta} = {\Phi}\ell_{\alpha}\ell_{\beta}\mathbf{\ell}\,,
\end{equation}
with
\begin{equation}
{\bf \ell} = \ell_{\alpha}{\vartheta}^{\alpha} \quad {\rm and} \quad
\ell^2: = g^{\alpha\beta}\ell_{\alpha}\ell_{\beta}=\ell_{\alpha}\ell^{\alpha}\,.
\end{equation}
For this ansatz one should compare Obukhov
\cite{Yuripp-waves,Yuripp-waves2006} who introduced plain fronted
waves in MAG, see also our considerations on the spin-$3$ solutions in
(\ref{solus}). 

Because of (\ref{ansatz_Q1}), the components of
the 1-form $Q_{\alpha\beta}$ become totally symmetric, i.e.,
\begin{equation}\label{totsym}
Q_{\g\alpha\beta} = Q_{(\g\alpha\beta )} =
{\Phi}\ell_{\g}\ell_{\alpha}\ell_{\beta}\, .
\end{equation}
{}From there on, we will put $n=4$. Because of (\ref{totsym}), the
irreducible pieces of the nonmetricity will simplify.  Together with
the one-forms
\begin{equation}
Q_{\alpha}{}^{\alpha} = 4Q = {\Phi}\ell^2{\bf \ell}\,,
\end{equation}
\begin{equation}\label{qslash*}
  {\qslash}_{\alpha\beta}=Q_{\alpha\beta}-Qg_{\alpha\beta}={\Phi}\left(
    \ell_{\alpha}\ell_{\beta}-\frac{1}{4}g_{\alpha\beta}\ell^2\right){\bf
    \ell}\,,
\end{equation}
\begin{equation}\label{Lamb}
{\Lambda}=\left( e^{\beta}\rfloor \,
  {\qslash}_{\alpha\beta}\right)\,
{\vartheta}^{\alpha}=\frac{3}{4}{\Phi}\ell^2{\bf \ell} = 3Q\,,
\end{equation}
and the two-form
\begin{equation}\label{qslash**}
P_{\alpha}:={\qslash}_{\alpha\beta}\wedge
{\vartheta}^{\beta}-\frac{1}{3}{\vartheta}_{\alpha}\wedge
{\Lambda} = 0\,,
\end{equation}
we find for the irreducible parts of the nonmetricity
\begin{eqnarray}\label{hurrah}
  ^{(1)}Q_{\alpha\beta} & = & {\Phi}\left(
    \ell_{\alpha}\ell_{\beta}-\frac{1}{6}\ell^2g_{\alpha\beta}\right){\bf \ell}-
 \frac{1}{3}{\Phi}\ell^2\ell_{(\a}{\vartheta}_{\b)} =
  \Phi\left(\ell_\a\ell_\b\ell_\g-\frac 12\ell^2g_{(\a\b}\ell_{\g)} 
  \right)\vt^\g\,,
  \cr & & \cr
  ^{(2)}Q_{\alpha\beta} & = &-\frac{2}{3}\,e_{(\a}\rfloor P_{\b)}=0\,, 
  \cr & & \cr ^{(3)}Q_{\alpha\beta} &
  = & \frac{1}{3}{\Phi}\ell^2\left(\ell_{(\a} {\vartheta}_{\b)} -
    \frac{1}{4}g_{\alpha\beta}{\bf \ell}\right)=
  \frac{1}{3}{\Phi}\ell^2\left( g_{\g(\a}\ell_{\b)}-\frac
      14g_{\a\b}\ell_\g
 \right)\vt^\g\,, \cr & & \cr
  ^{(4)}Q_{\alpha\beta} & = &
  \frac{1}{4}{\Phi}\ell^2g_{\alpha\beta}{\bf \ell}\,.
\end{eqnarray}
Since $^{(1)}Q_{\a\b}\ne 0$, the ansatz (\ref{ansatz_Q1}) may carry
{\it genuine} spin three. This is consistent with (\ref{solus}) and
(\ref{kkk}) and with the fact that the helicity-3 plane-wave solutions
obey Bargmann--Wigner equations for spin three. Observe that the main
spin-2 contribution, mediated by the tensor part
$^{(2)}Q_{\alpha\beta}$, vanishes identically. By using
(\ref{qslash*}) to (\ref{qslash**}) and the vanishing of the torsion,
$T^\a=0$, it is possible to show that $^{(2)}Z_{\a\b}=0$ and
$^{(3)}Z_{\a\b}\sim d\Lambda$, see Appendix \ref{zeroB},
Eq.(\ref{result2}).

Furthermore, we will need the Hodge duals of $^{(1)}Q_{\alpha\beta}$
and of the other irreducible pieces. Here the $\eta$-basis
(\ref{etabasis}) is very convenient. The $^{(1)}Q_{\alpha\beta}$, as
expressed in terms of $\vt^\g$, can be easily hodged:
\begin{equation}\label{*1Q}
^{\star(1)}Q_{\alpha\beta} =
  \Phi\left(\ell_\a\ell_\b\ell_\g-\frac 12\ell^2g_{(\a\b}\ell_{\g)} 
  \right)\eta^\g\,. 
\end{equation} 
It works for the other pieces analogously. If we recall
$\vt^\a\wedge\eta^\b=g^{\a\b}\eta$, see \cite{PRs}, then, by
straightforward algebra, we find
\begin{eqnarray}
 ^{(1)}Q_{\alpha\beta}\wedge{} ^{\star(1)}Q^{\alpha\beta} &=&
\frac 12\Phi^2\ell^6\eta\,,\cr
 ^{(3)}Q_{\alpha\beta}\wedge{} ^{\star(3)}Q^{\alpha\beta} &=&
\frac 14\Phi^2\ell^6\eta\,,\cr
 ^{(4)}Q_{\alpha\beta}\wedge{} ^{\star(4)}Q^{\alpha\beta} &=&
\frac 14\Phi^2\ell^6\eta\,.\label{Q^2}
\end{eqnarray}
We transvect (\ref{*1Q}) with $\ell^{\beta}$ and find
\begin{equation}
  ^{\star (1)}Q_{\alpha\beta}\ell^{\beta} = \frac{2}{3}{\Phi}\ell^2\left(
    \ell_{\a}\ell_{\b}-\frac{1}{4}\ell^2g_{\alpha\b}\right){\eta}^{\b}\,.
\end{equation} 
Additionally, a couple of relations for the nonmetricity as
multiplied by $\eta^{\a\b}$ will be needed for simplifying
the field equations. We use the ansatz (\ref{ansatz_Q1}) and the
properties of the ${\eta}$-bases, cf.\ \cite{PRs},
\begin{eqnarray}\label{Q_eta}
  Q_{\g\a}\wedge {\eta}^{\g}{}_{\beta} & = &
  {\Phi}\ell_{\alpha}\ell_{\beta}\ell_{\g}{\eta}^{\g}-{\Phi}\ell^{2}
  \ell_{\alpha}\eta_{\beta}\,
  , \cr Q\wedge {\eta}_{\alpha\beta} & = &
  -\frac{1}{2}{\Phi}\ell^2\ell_{[\alpha}{\eta}_{\beta]}\, , \cr
  Q_{\g[\a}\wedge {\eta}^{\g}{}_{\beta ]} & = &
  -{\Phi}\ell^{2}\ell_{[\alpha}{\eta}_{\beta ]}=2Q\wedge
  {\eta}_{\alpha\beta} \, , \cr Q_{\g(\a}\wedge
  {\eta}^{\g}{}_{\beta )} &  = &
  {\Phi}\ell_{\alpha}\ell_{\beta}\ell_{\g}{\eta}^{\g}-{\Phi}\ell^{2}
  \ell_{(\alpha}\eta_{\beta)}\,.
\end{eqnarray}

Some consequences of the ansatz (\ref{ansatz_Q1}) that we will use
over and over again are the following relations:
\begin{eqnarray}\label{ans}
  &&Q_{\a\b}=\Phi\ell_\a\ell_\b\ell\,,\quad 
  \qslash_{\a\b}\sim\ell\,,\\ \label{sy}
  &&  e_\g\rfloor Q_{\a\b}=Q_{\g\a\b}=Q_{(\g\a\b)}\,,\qquad
  e_{[\g}\rfloor Q_{\a]\b}=Q_{[\g\a]\b}=0\,,\\
\label{connansatz}
&&  \Gamma_{\a\b}=\widetilde{\Gamma}_{\a\b}-e_{[\a}\rfloor T_{\b]}+
\frac 12(e_\a\rfloor e_\b\rfloor T_\g)\vt^\g+\frac 12Q_{\a\b}\,,\\
\label{QLam}
&&Q=\frac 13\Lambda=\frac 14\Phi\ell^2\ell\,,\qquad Q\wedge \Lambda=0\\
\label{P2Q}
&& P_\a= 0\quad \qquad{}^{(2)}Q_{\a\b}=0\,.
\end{eqnarray}

Furthermore, we assume for the rest of Sec.IV, similar as Boulanger
and Kirsch \cite{Kirsch:2005st,Boulanger:2006tg}, that the {\it
  torsion vanishes:}
\begin{equation}\label{vanT}
T^\a=0\,.
\end{equation}
This implies that the connection (\ref{connansatz}) reduces to
\begin{equation}\label{Gamma_1}
{\Gamma}_{\alpha\beta} =
\widetilde{{\Gamma}}_{\alpha\beta}+\frac{1}{2}Q_{\alpha\beta}\quad{\rm
  or}\quad N_{\alpha\beta}=\frac{1}{2}Q_{\alpha\beta}\, .
\end{equation}
Connections of this type have been studied in a different context
by Baekler et al.\ \cite{MAGIII,Magaxi}.
%
\subsection{A pure $^{(1)}Q_{\a\b}$ square Lagrangian}
\label{PURE1Q}
%
In order to understand a propagating connection, we consider first as
a very special and degenerate case of (\ref{QMA}) the simple field
Lagrangian
\begin{equation}\label{lag_Q1}
V_{^{(1)}Q^{2}} = \frac{b_{1}}{2\kappa}\,Q_{\alpha\beta}\wedge \,
^{{\star}(1)}Q^{\alpha\beta}\,.
\end{equation}
The corresponding excitations (\ref{M-excit}),
(\ref{Ha-excit}), and (\ref{Hab-excit}) turn out to be
\begin{equation}{\label{excitations2}}
  M^{\alpha\beta} =
  -\frac{2}{\kappa}b_{1}\,^{{\star}(1)}Q^{\alpha\beta}\,,\quad
  H_{\alpha}
  =  0\, , \quad  H^{\alpha}{}_{\beta} =  0 \,,
\end{equation}
and the gauge currents (\ref{zerothx}), (\ref{firstx}), and
(\ref{secondx}) read
\begin{eqnarray}\label{zer*}
  m^{\alpha\beta} & = &
  {\vartheta}^{(\alpha}\wedge E^{\beta )} +
  Q^{(\beta}{}_{\gamma}\wedge M^{\alpha ) \gamma}\,,\\ \label{fir*}
  E_{\alpha} & = & e_{\alpha}\rfloor \, V_{^{(1)}Q^2} + \frac{1}{2}\left(
    e_{\alpha}\rfloor \, Q_{\beta\gamma}\right) M^{\beta\gamma}\, ,
  \\ E^{\alpha}{}_{\beta} & = & -g_{\beta\gamma}M^{\alpha\gamma}=
  \frac{2}{\kappa}b_{1}g_{\beta\gamma}\,
  ^{{\star}(1)}Q^{\alpha\g}\,.\label{sec*}
\end{eqnarray}
Then the source-free field equations (\ref{zeroth}), (\ref{first}),
and (\ref{second}) reduce to
\begin{eqnarray}\label{zer}
  DM^{\alpha\beta} - m^{\alpha\beta} & = & 0\, , \\\label{fir}
  E_{\alpha} & = & 0\, , \\
  E^{\alpha}{}_{\beta} & = & 0\,.\label{sec}
\end{eqnarray}

This is a rather trivial case. Because of (\ref{sec}) and
(\ref{sec*}), we have
\begin{equation}\label{vanQ}
  ^{(1)}Q^{\alpha\beta}=0 \,.
\end{equation}
Thus, also $M^{\a\b}=0$, and the field equations are identically
fulfilled. Consequently, the source-free field equations corresponding
to the purely quadratic Lagrangian (\ref{lag_Q1}) do not allow for
propagating spin-3 fields. Our ansatz (\ref{ans}) was not needed
in order to achieve this result.

All this seems hardly surprizing. However, we have to be aware that
$Q_{\a\b}=-Dg_{\a\b}$ is itself a field strength. Hence a check of the
triviality of the Lagrangian (\ref{lag_Q1}) was desirable.

\subsection{Adding a Hilbert--Einstein type term}
\label{+HilbertE}
%
Let us augment the Lagrangian (\ref{lag_Q1}) by a curvature piece, the
simplest one being the curvature scalar, and a cosmological term. In
this case the langrangian assumes the form
\begin{equation}\label{lag_RQ1}
V_{\rm R+{}^{(1)}Q^2} = -\frac{a_{0}}{2\kappa}R_{\alpha\beta}\wedge
{\eta}^{\alpha\beta}-\frac{{\lambda}_{0}}{\kappa}{\eta} +
\frac{b_{1}}{2\kappa}\,Q_{\alpha\beta}\wedge \,
^{{\star}(1)}Q^{\alpha\beta}\,.
\end{equation}
Besides the gravitational constant $\kappa$ and the cosmological
constant $\lambda_0$, we have $a_0=+1$ or $=0$ (for switching on and
off) and $b_1=\;arbitrary$ as dimensionless coupling constants.  For
this particular Lagrangian, the excitations turn out to be
\begin{equation}{\label{excitations1}}
  M^{\alpha\beta}  =
  -\frac{2}{\kappa}b_{1}\,^{{\star}(1)}Q^{\alpha\beta}\,,\quad
  H_{\alpha} =  0\, , \quad  H^{\alpha}{}_{\beta}  =
  \frac{a_{0}}{2\kappa}{\eta}^{\alpha}{}_{\beta} \,.
\end{equation}

Substitution of (\ref{excitations1}) into the second sourcefree field
equation yields the algebraic relation
\begin{equation}\label{second_1}
\frac{a_{0}}{2\kappa}\left( Q^{\alpha\g}\wedge {\eta}_{\g\beta}
- 2Q\wedge {\eta}^{\alpha}{}_{\beta} + T^{\g}\wedge
{\eta}^{\alpha}{}_{\beta\g}\right)
-\frac{2}{\kappa}b_{1}\,^{\star(1)}Q^{\alpha}{}_{\beta} = 0\,.
\end{equation}
We now substitute the ansatz (\ref{ans}), (\ref{*1Q}), and
(\ref{vanT}) into (\ref{second_1}):
\begin{equation}\label{second_2}
  \frac{a_{0}}{2}\left( {\Phi}\ell_{\alpha}\ell_{\g}{\bf \ell}\wedge
    {\eta}^\g{}_\beta-\frac{1}{2}{\Phi}\ell^2{\bf \ell}\wedge
    {\eta}_{\alpha\beta}\right)-2b_{1}\left[
    {\Phi}\ell_{\alpha}\ell_{\beta}\ell_{\g}{\eta}^{\g}
    -\frac{1}{6}{\Phi}\ell^2\left(
      g_{\alpha\beta}\ell_{\g}{\eta}^{\g}+\ell_{\alpha}{\eta}_{\beta}
      +\ell_{\beta}{\eta}_{\alpha}
    \right) \right] = 0\,.
\end{equation}
Transvection with $\ell^{\beta}$ yields
\begin{equation}
  \left(
    -\frac{1}{4}a_{0}+\frac{1}{3}b_{1}\right){\Phi}\ell^4{\eta}_{\alpha}+
  \left(
    \frac{1}{4}a_{0}-\frac{4}{3}b_{1}\right){\Phi}\ell^2\ell_{\alpha}
  \ell_{\g}{\eta}^{\g}  = 0\,.
\end{equation}
The second field equation (\ref{second_1}) is only fulfilled by the
choice
\begin{equation}\label{l_2}
\ell^2 =0\,.
\end{equation}
We substituting this into (\ref{second_2}) and obtain
\begin{equation}
  \left(\frac{a_{0}}{2}-2b_{1}\right){\Phi}\ell_{\alpha}
  \ell_{\beta}\ell_{\g}{\eta}^{\g}= 0\,.
\end{equation}
The only choice for non-trivial field configurations is
\begin{equation}
b_{1} = \frac{a_{0}}{4}\quad {\rm and}\quad \ell^2=0\,.
\end{equation}

What about the first field equation? Because of (\ref{l_2}), the Hodge
dual of $^{\star(1)}Q_{\alpha\beta}$ reduces to
\begin{equation}\label{Q1_star*}
  ^{\star(1)}Q_{\alpha\beta} =
  {\Phi}\ell_{\alpha}\ell_{\beta}\ell_{\g}{\eta}^{\g}\,,\quad{\rm
    and}\quad
  ^{(1)}Q_{\alpha\beta}\wedge{} ^{\star(1)}Q^{\alpha\beta}=0\,.
\end{equation}
To simplify the gauge current $E_{\alpha}$ in (\ref{firstx}), we need
information about $(e_{\alpha}\rfloor\,
Q_{\beta\gamma})M^{\beta\gamma}$. Because of (\ref{l_2}), this can be
shown to be identically zero. Collecting our results, the first
sourcefree field equation (\ref{first}) reduces to
\begin{equation}
E_{\alpha}=e_{\alpha}\rfloor\, V_{\rm R+{}^{(1)}Q^2} +
\frac{a_{0}}{2\kappa}\left(e_{\alpha}\rfloor\,
R_{\beta}{}^{\gamma}\right)\wedge {\eta}^{\beta}{}_{\gamma} = 0
\end{equation}
or, with $\lambda=\lambda_0/a_0$, to
\begin{equation}\label{Einstein}
  G_{\alpha} + {\lambda}{\eta}_{\alpha} = 0\,,
\end{equation}
where $G_{\alpha}$ is the Einstein three-form (\ref{Einstein3}) that
will determine the one-form ${\ell}$ and the scalar field ${\Phi}$.

We can decompose the first field equation (\ref{Einstein}) into
Riemannian and post-Riemannian pieces. For this purpose we start with
the antisymmetric part of (\ref{curv}),
\begin{equation}\label{curv'}
  W^{\alpha\beta} = R^{[\alpha\beta]}={\widetilde R}^{\alpha\beta}+
  {\widetilde D}N^{[\alpha\beta]} - N^{[\alpha|\g|}\wedge
  N_{\g}{}^{\beta]}\,
\end{equation}
in which (\ref{Gamma_1}) is substituted:
\begin{equation}\label{curv''}
  W^{\alpha\beta} ={\widetilde R}^{\alpha\beta}
  -\frac 14 Q^{[\alpha|\g|}\wedge
  Q_{\g}{}^{\beta]}\,.
\end{equation}
The last two terms vanish since $ Q^{\a\g} \wedge Q_\g{}^\b
={\Phi}^2\ell^\a\ell^{\g}\ell_{\g}\ell^\b{\bf \ell}\wedge
\mathbf{\ell} = 0$. Thus,
\begin{equation}\label{Einstein'}
  G_\a=\frac 12\eta_{\a\b\g}\wedge W^{\b\g}=\frac
  12\eta_{\a\b\g}\wedge {\widetilde R}^{\b\g}=\widetilde{G}_\a\,.
\end{equation}
Hence our field equation reads $\widetilde{G}_{\alpha} +
{\lambda}{\eta}_{\alpha} = 0$ or, in components of the (Riemannian)
Einstein tensor,
\begin{equation}\label{V4_Einstein}
{\widetilde{G}}_{\alpha\beta} + {\lambda}g_{\alpha\beta} = 0\,.
\end{equation}

Observe that (\ref{Einstein}) to leading order yields
\begin{equation}\label{leading}
D\,^{(1)}Q_{\alpha\beta} + {\rm nonlinear\,\, terms} = 0\,.
\end{equation}
To separate the maximal spin content $s=3$ of the connection, we have
to take the totally symmetric part of (\ref{Gamma_1}):
\begin{equation} {\Gamma}_{(\g\alpha\beta )} =
  \frac{1}{2}Q_{(\g\alpha\beta )} +\frac 12\partial_{(\g}g_{\a\b)} =
  \frac{1}{2}{\Phi}\ell_{\g}\ell_{\alpha}\ell_{\beta}+\frac
  12\partial_{(\g}g_{\a\b)}\stackrel{*}{=}
  \frac{1}{2}{\Phi}\ell_{\g}\ell_{\alpha}\ell_{\beta}\, .
\end{equation}
The star denotes the choice of an orthonormal frame. However, as we
have seen, these terms drop out from (\ref{Einstein}) and only the
Riemannian counterpart (\ref{V4_Einstein}) is left.

Anyway, any solution of Einstein's field equation with cosmological
constant will generate (massless) fields with spin-3 content in the
framework of MAG.  It remains to be seen whether this fact is of
physical relevance. In any case, it shows that higher-spin fields can
be constructed {}from the field equations of MAG.
Transvection of (\ref{V4_Einstein}) with $\ell^{\beta}$ yields
\begin{equation}
{\widetilde{G}}_{\alpha}{}^{\beta}\ell_{\beta}={\lambda}\ell_{\alpha}\,.
\end{equation}
This is an eigenvalue equation for the eigenvector $\ell^{\alpha}$, and
the cosmological constant ${\lambda}$ is the corresponding eigenvalue
of the (Riemannian) Einstein-tensor.
%
%
\subsection{Still more $Q_{\a\b}$ square terms added for spin 3 
fields with $\ell^2\neq 0$}
\label{allQ2}
%
The gravitational sector allows also for spin 3 modes with $\ell^2\neq
0$. We call them tentatively {\it massive} modes since we interpret
$\ell$ as wave covector. To support the connection $\Gamma_\a{}^\b$ to
carry massive modes of this type, the Lagrangian (\ref{lag_RQ1}) has
to be extended in order to include, besides $^{(1)}Q_{\alpha\beta}$,
also the other irreducible pieces of the nonmetricity. These
contributions will induce massive spin 3 parts in the connection. As a
suitable Lagrangian with this property we choose
\begin{equation} V_{{\rm R+Q^{2}}} =
-\frac{a_{0}}{2\kappa}R^{\alpha\beta}\wedge {\eta}_{\alpha\beta} -
\frac{{\lambda}_{0}}{\kappa}{\eta}+
\frac{1}{2\kappa}Q_{\alpha\beta}\wedge \sum\limits_{I=1}^{4}
b_{I}\,^{\star(I)}Q^{\alpha\beta}\, .
\end{equation}
The corresponding excitations are
\begin{equation}\label{mom}
  M^{\a\b}=-\frac{2}{\kappa}\sum\limits_{I=1}^{4}
  b_{I}\,^{\star(I)}Q^{\alpha\beta}\,,\quad H_{\alpha}=0\, , \quad
  H^{\alpha}{}_{\beta}=\frac{a_{0}}{2\kappa}{\eta}^{\alpha}{}_{\beta}\,.
\end{equation}
Accordingly, the second field equation (\ref{second}) [with
(\ref{secondx})] is again algebraic:
\begin{equation}\label{second_3}
\frac{a_{0}}{2\kappa}\left( Q_\alpha{}^\g\wedge {\eta}_{\g\beta}
- 2Q\wedge {\eta}_{\alpha\beta} + T^{\g}\wedge
{\eta}_{\alpha\beta\g}\right) +M_{\a\b} = 0\,.
\end{equation}
Its trace, its symmetric, and its antisymmetric pieces read,
respectively,
\begin{eqnarray}\label{trace2nd}
M_\g{}^\g=0\qquad{\rm or}\qquad b_4Q &=& 0\,,\\
\label{symm2nd}
\frac{a_{0}}{2\kappa}Q_{(\alpha}{}^\g\wedge
{\eta}_{|\g|\beta)}+M_{\a\b}&=&0 \,,\\
  \frac{a_{0}}{2\kappa}\left( Q_{[\alpha}{}^\g\wedge
    {\eta}_{|\g|\beta]} - 2Q\wedge {\eta}_{\alpha\beta}
    + T^{\g}\wedge{\eta}_{\alpha\beta\g}\right)&=&0\,.\label{antisymm2nd}
\end{eqnarray}

In the case of vanishing torsion $T^\a=0$ and the application of
(\ref{ans}) in combination with (\ref{Q_eta}), Eq.\
(\ref{antisymm2nd}) vanishes identically and is thus fulfilled, and
the symmetric part (\ref{symm2nd}) becomes
\begin{equation}\label{symm2nd'}
\frac{a_{0}}{2\kappa} {\Phi}\left(\ell_{\alpha}\ell_{\beta}\ell_{\g}{\eta}^{\g}
-\ell^2\ell_{(\a}\eta_{\b)}\right)+M_{\a\b}=0\,.
\end{equation}
With the ansatz (\ref{ans}), we find for $M_{\a\b}$ in
(\ref{mom})
\begin{eqnarray}\label{M_ab}
 M_{\alpha\beta} & = &-\frac{2\Phi}{\kappa}\left[
b_{1}\ell_{\alpha}\ell_{\beta}\ell_{\g} -\frac{1}{12}
\left(2b_{1}+b_{3}-3b_{4}\right)\ell^{2}
g_{\alpha\beta}\ell_{\g} \right.\cr & &\left.\hspace{60pt}
-\frac{1}{3}\left(b_{1}-b_{3}\right)
\ell^{2}g_{\g(\a}\ell_{\b)}\right]\eta^\g\, .
\end{eqnarray}
We substitute this into (\ref{symm2nd'}) and find a new form of the
symmetric part of the second field equation:
\begin{eqnarray}\label{symm2nd''}
 && \frac{\Phi}{6\kappa}\left[3
    (a_0-4b_{1})\ell_{\alpha}\ell_{\beta}\ell_{\g} +
    \left(2b_{1}+b_{3}-3b_{4}\right)\ell^{2}
    g_{\alpha\beta}\ell_{\g}\right.\cr
   &&\left.\hspace{50pt} + \left(-3a_0+4b_{1}-4b_{3}\right)
    \ell^{2}g_{\g(\a}\ell_{\b)}\right]\eta^\g=0\, .
\end{eqnarray}
The $b_4$-term in this equation is $\sim b_4{}^\star Q$. Because of
(\ref{trace2nd}), it drops out. We transvect this equation first with
$\ell^\b$,
\begin{eqnarray}\label{symm2nd'''}
\left[(\frac 32a_0-8b_1-b_3)\ell^2\ell_\a\ell_\g+
(-\frac 32a_0+2b_1-2b_3)\ell^4 g_{\a\g}\right]\eta^\g=0\, ,
\end{eqnarray}
and subsequently with $\ell^\a$,
\begin{eqnarray}\label{symm2nd''''}
-3(2b_1+b_3)\ell^4\ell_\g\eta^\g=0\,.
\end{eqnarray}
Provided $\ell^2\ne 0$, we have from (\ref{trace2nd}) and from
(\ref{symm2nd''''}) the relations $ b_4=0$ and $b_3=-2b_1$,
respectively. If we substitute the latter into (\ref{symm2nd'''}), we
have finally
\begin{equation}\label{parameterset_3}
b_{1}=\frac{1}{4}a_{0}\, , \quad b_{3}=-\frac{1}{2}a_{0}\,,
\quad b_4=0\,, \quad{\rm all\;for}\;\ell^{2}\neq 0\,.
\end{equation}

For a reformulation of the first field equation (\ref{first}) [with
(\ref{firstx})], 
\begin{equation}\label{fi}
  E_{\alpha} =e_{\alpha}\rfloor V_{{\rm R+Q^{2}}}
  + (e_{\alpha}\rfloor R_{\beta}{}^{\gamma})\wedge H^{\beta}{}_{\gamma} +
  {1\over2}(e_{\alpha}\rfloor Q_{\beta\gamma})\, M^{\beta\gamma}=0\,,
\end{equation}
we use (as part of $ V_{{\rm R+Q^{2}}}$)
\begin{equation}
  Q_{\alpha\beta}\wedge \sum\limits_{I=1}^{4}
  b_{I}\,^{\star(I)}Q^{\alpha\beta} =\left(\frac 12b_{1}+\frac 14b_{3}
    +\frac 14b_{4}\right){\Phi}^{2}\ell^{6}{\eta}
\end{equation}
and
\begin{equation}\label{M_Q}
  \frac{1}{2}Q_{\alpha\beta\gamma}M^{\beta\gamma}=-\frac{1}{\kappa}
  \left(\frac{1}{2}b_{1}+\frac{1}{4}b_{3}
    +\frac{1}{4}b_{4}\right){\Phi}^2\ell^{4}\ell_{\alpha}
  \ell_{\b}{\eta}^{\b}\, .
\end{equation}
If we collect our results, (\ref{fi}) can be written as
\begin{eqnarray}
  E_{\alpha}&=&-\frac{a_{0}}{\kappa}G_{\alpha\b}{\eta}^{\b}
  -\frac{{\lambda}_{0}}{\kappa}\eta_\a+
  \frac{1}{2\kappa}\left(\frac{1}{2}b_{1}+\frac{1}{4}b_{3}
    +\frac{1}{4}b_{4}\right){\Phi}^{2}\ell^{6}
  {\eta}_{\alpha}\cr & & \cr & &
  -\frac{1}{\kappa}\left(\frac{1}{2}b_{1}+\frac{1}{4}b_{3}
    +\frac{1}{4}b_{4}\right)
  {\Phi}^{2}\ell^{4}\ell_{\alpha}\ell_{\b}{\eta}^{\b}=0\, .
\end{eqnarray}
Eventually, the first field equation reads
\begin{equation}
  \frac{a_{0}}{\kappa}\left(
    G_{\alpha\b }+\frac{{\lambda}_{0}}{a_{0}}g_{\alpha\b }\right){\eta}^{\b }
  +\frac{1}{\kappa}\left(\frac{1}{2}b_{1}+\frac{1}{4}b_{3}
    +\frac{1}{4}b_{4}\right)  {\Phi}^2\ell^{4}\left(
    \ell_{\alpha}\ell_{\b }-\frac{1}{2}\ell^{2}g_{\alpha\b }\right)
  {\eta}^{\b }=0\,.
\end{equation}

Using the parameter set (\ref{parameterset_3}), the expression
containing $b_1$ etc.\ collapses to zero and we end up with an
Einstein-type vacuum equation
\begin{equation}
  G_{\alpha\beta}({\Gamma})+{\lambda}g_{\alpha\beta}=0\,,
\end{equation}
where we put again ${\lambda}={\lambda}_{0}/a_{0}$. As in the last
subsection, this equation, using our ansatz (\ref{ans}) and
(\ref{vanT}), reduces to the Einstein equation in Riemannian
spacetime:
\begin{equation}
  \widetilde{G}_{\alpha\beta}+{\lambda}g_{\alpha\beta}=0\,.
\end{equation}

In our context, the Einstein three-form $G_\a(\Gamma)$ equals the
Riemannian one $G_\a(\widetilde{\Gamma})\equiv\widetilde{G}_\a$. There
is a general underlying pattern. If a connection is deformed by means
of an additive one-form $A_\a{}^\b$ according to
$\overline{\Gamma}_\a{}^\b={\Gamma}_\a{}^\b+A_\a{}^\b$, then the
curvature tensor responds with 
\begin{eqnarray}\overline{R}_{\alpha}{}^{\beta} &=& R_{\alpha}{}^{\beta} +
  DA_{\alpha}{}^{\beta} - A_{\alpha}{}^{\g}\wedge A_{\g}{}^{\beta}\,.
\end{eqnarray}
In the special case of a projective transformation with
$A_\a{}^\b=\delta_\a^\b P$, we have (see \cite{PRs,Macias:1995hf})
\begin{equation}
  \stackrel{\rm proj.}{R_\a{}^\b}=R_\a{}^\b+\d_\a^\b\,dP\,.
\end{equation}
Thus, 
\begin{equation}\stackrel{\rm proj.}{W_{\a\b}}\,:=\,\stackrel{\rm
    proj.}{R_{[\a\b]}}\,=R_{[\a\b]}=W_{\a\b}\quad{\rm and}\quad 
  \stackrel{\rm proj.}{G_\a}=G_\a\,.
\end{equation}
The Einstein three-form is invariant under projective transformations.
Therefore, a gravitational Lagrangian in MAG cannot consist of a
Hilbert--Einstein type term alone. It has to carry additional terms.

The connection of our ansatz (\ref{Gamma_1}), namely
${\Gamma}_{\alpha\beta}={\widetilde{\Gamma}}_{\alpha\beta}
+\displaystyle\frac{1}{2}Q_{\alpha\beta}$, transforms the curvature
according to
\begin{equation}
  R_{\a\b}(\Gamma)=\widetilde{R}_{\a\b}+\frac 12\widetilde{D} Q_{\a\b}-
  \frac 14Q_\a{}^\g\wedge Q_{\b\g}\,.
\end{equation}
Consequently,
\begin{equation}
  W_{\a\b}(\Gamma)=\widetilde{W}_{\a\b}- \frac 14Q_\a{}^\g\wedge Q_{\b\g}\,,
\end{equation}
since the last term is antisymmetric in $\a$ and $\b$. In turn,
\begin{equation}
G_\a(\Gamma)=\widetilde{G}_\a+\frac 12
\eta_{\a\b\g}Q^\b{}_\delta\wedge Q^{\g\d}\,.
\end{equation}
However, in accordance with our ansatz (\ref{ans}), the Q-square term
vanishes:
\begin{equation}
  G_\a(\Gamma)=\widetilde{G}_\a\,,\qquad  W_{\a\b}(\Gamma)=
\widetilde{W}_{\a\b}\,.
\end{equation}
%
\subsection{A quadratic Lagrangian with pure strain curvature}
\label{YM}
%
In reminiscence of the Fronsdal Lagrangian, let us investigate a
gravitational gauge model in the framework of {\bf MAG} with a field
Lagrangian quadratic in the (symmetric) strain curvature,\footnote{A
  Lagrangian quadratic in the rotational curvature of the type
  $W^{\a\b}\wedge{}^\star W_{\a\b}$ would not have a propagating
  $^{(1)}Q_{\a\b}$ piece. This can be seen as follows: The third term
  on the right-hand-side of (\ref{N}) selects all the pieces of
  $Q_{\a\b}$, except $^{(1)}Q_{\a\b}$. The fourth term will not
  contribute to give a kinetic term $d{}^{(1)}Q^{\a\b}\wedge{}^\star
  d{}^{(1)}Q_{\a\b}$ via $W^{\a\b}\wedge{}^\star W_{\a\b}$ because of
  its symmetries.  Therefore, only the third term of (\ref{N}) has a
  chance to contribute a kinetic term $dQ^{\a\b}\wedge{}^\star
  dQ_{\a\b}$; but in the third term $^{(1)}Q_{\a\b}$ dropped out.}
i.e., we will concentrate on the field Lagrangian
\begin{equation}\label{V_B}
V_{\rm Z^2} = -\frac{1}{2\rho}R^{\alpha\beta}\wedge \sum\limits_{I=1}^{5}
z_{I}\, ^{\star (I)}Z_{\alpha\beta}\, .
\end{equation}
Incidentally, such Lagrangians may be also interesting in cosmology,
see Puetzfeld \cite{Puetzfeld2004a,Puetzfeld2004c}. The excitations
belonging to the Lagrangian (\ref{V_B}) turn out to be
\begin{equation}
 M^{\alpha\beta}=0\, , \quad H_{\alpha}=0\, ,\quad
 H^{\alpha}{}_{\beta}=\frac{1}{\rho}\sum\limits_{I=1}^5 z_{I}\, ^{\star (I)}
Z^{\alpha}{}_{\beta}\, .
\end{equation}
Note that $H^{\a\b}$ is symmetric in $\a$ and $\b$. The source-free
field equations (\ref{first}), (\ref{second}) reduce to
\begin{eqnarray}\label{Z_field}
 e_{\alpha}\rfloor V_{\rm Z^2} + \frac{1}{{\rho}}\left(e_{\alpha}\rfloor
 Z_{\beta}{}^{\gamma}\right)\wedge
\sum\limits_{I=1}^5 z_{I}\, ^{\star (I)} Z^{\beta}{}_{\gamma}
 & = & 0\, , \\ & & \cr
 D\left(\sum\limits_{I=1}^5 z_{I}\, ^{\star (I)}
Z^{\alpha}{}_{\beta}\right) & = & 0\, .\label{Z_field2}
\end{eqnarray}
The trace of the second field equation (\ref{Z_field2}) yields
\begin{equation}\label{z4_dstar_dQ}
 2z_{4}\,d\,^{\star}\!dQ = 0\, 
\end{equation}
and from its antisymmetric piece only
\begin{equation}\label{antisymmetricpiece}
Q_{\g[\alpha|}\sum\limits_{I=1}^{5} z_{I}\, ^{\star (I)}
Z_{|\beta]}{}^{\gamma}=0
\end{equation}
is left over.

In order to get some insight into the possible solution classes, we
will distinguish between $\ell^2=0$ and  $\ell^2\ne 0$

\subsubsection{Solutions with ${\ell}^2= 0$}\label{YMzero}

Let us first recall {}from (\ref{QLam}) that the Weyl covector $Q$,
for ${\ell}^2=0$, vanishes identically. Hence
$^{(4)}Z_{\alpha\beta}=0$. Again with our ansatz, according to
(\ref{result2}), we have $^{(2)}Z_{\alpha\beta}=0$ and
$^{(3)}Z_{\alpha\beta}\sim d\Lambda$. However, $\Lambda\sim Q$, see
(\ref{QLam}). Accordingly,
\begin{equation}
^{(2)}Z_{\alpha\beta}=\,^{(3)}Z_{\alpha\beta}=\, ^{(4)}Z_{\alpha\beta} = 0\, .
\end{equation}

To find solutions of the field equations (\ref{Z_field}), we will make
use of the {\it Kerr--Schild ansatz} for the metric, cf.\
\cite{GurGur}, which will be expressed in terms of a null-tetrad
according to
\begin{equation}
  g=g_{\alpha\beta}{\vartheta}^{\alpha}\otimes {\vartheta}^{\beta}=
  {\vartheta}^{0}\otimes{\vartheta}^{1}+{\vartheta}^{1}\otimes{\vartheta}^{0}
  -{\vartheta}^{2}\otimes{\vartheta}^{3}-{\vartheta}^{3}\otimes{\vartheta}^{2}\,,
\end{equation}
that is, the anholonomic components of the (local) metric are given by
\begin{equation}
g_{\alpha\beta}=\pmatrix{0&1&0&0\cr 1&0&0&0\cr 0&0&0&-1\cr
0&0&-1&0}\,.
\end{equation}
We will introduce a set of coordinates $(\zeta\,, \oz\, , u,v)$ and
choose the coframe
\begin{eqnarray}\label{coframe_0}
{\vartheta}^{0} & = & d{\zeta}\, , \cr {\vartheta}^{1} & = &
d{\oz}\, , \cr {\vartheta}^{2} & = & du\, , \cr {\vartheta}^{3} &
= & dv + V{\vartheta}^{2}\, .
\end{eqnarray}
Then, the metric assumes the form
\begin{equation}
g = 2(d{\zeta}d{\oz}-dudv)-2V({\zeta}\, , {\oz}\, , u)du^2 \,,
\end{equation}
which will generate a class of pp-waves, inter alia, cf.\
\cite{GursesHalil,Yuripp-waves,Yuripp-waves2006}.

The key point now is to identify the propagation vector ${\ell}$ of
the spin 3-field with that of the Kerr--Schild ansatz, i.e., we will
choose for the {\em propagating} {\tt trinom}
\begin{equation}\label{ell}
{\ell} = V({\zeta}\, , {\oz}\, , u)du\,,
\end{equation}
with the further property
\begin{equation}
{\ell}\wedge d{\ell} =  0\, .
\end{equation}
[In classical general relativity the components of ${\ell}^{\rm KS}$
are chosen to be ${\ell}_{\alpha}^{\rm KS}=(0,0,1,0)$.] Hence, in the
massless case, i.e., ${\ell}^2=0$, it would be of advantage to
re-scale the function ${\Phi}$ according to
\begin{equation}\label{Phi_rescale}
{\Phi}\rightarrow {\hat{\Phi}}/V\,,\quad{\rm with}\quad
{\hat\Phi}={\hat\Phi}(\zeta\, , \oz,u)\,.
\end{equation}
This rescaling introduces some redundancy. However, it is very
convenient when one searches for exact solutions of MAG. Then, one can
take, e.g., for $V$ an exact solution of Einstein's theory (in
Riemannian spacetime), but still has $\hat\Phi$ as a separate field for
fulfilling the field equations of MAG.

We insert (\ref{ans}) and (\ref{coframe_0}), together with
(\ref{Phi_rescale}), into the first field equation (\ref{Z_field}). It
is fulfilled identically for arbitrary parameter values of $z_{I}$.
The second field equation (\ref{Z_field2}) yields just one equation
for the determination of the functions $V$ and ${\hat\Phi}$,
\begin{eqnarray}\label{dgl_V}
  0 & = & z_{1}\left( {\hat{\Phi}}_{\zeta\oz}V^2 +
    2{ \hat{\Phi}}_{\zeta}V_{\oz}V+2{ \hat{\Phi}}_{\oz}V_{\zeta}V
    +2V_{\zeta\oz}{\hat {\Phi}}V+
    2V_{\zeta}V_{\oz}{\hat {\Phi}}\right)\cr & = &
  z_{1}\left({\hat {\Phi}}V^2\right)_{\zeta\oz}\,.
\end{eqnarray}
Incidentally, the choice $V=1$, that is ${\ell}=du$, would lead to
$z_{1}{\hat\Phi}_{\zeta\oz}=0\,$. Observe that in this case the
corresponding metric $g$ alone represents a flat spacetime whereas the
pair $\{g\,, {\hat\Phi}\}$ yields a non-flat solution of MAG. This is an
example that the re-scaling in (\ref{Phi_rescale}) pays dividends.

Substitution of the nonmetricity (\ref{ans}) and the coframe
(\ref{coframe_0}) together with the condition of vanishing torsion
yields for the massless case (${\ell}^{2}=0$)
\begin{equation}
^{(1)}W^{\alpha\beta}\neq 0\, , \quad ^{(4)}W^{\alpha\beta}\neq
0\, , \quad {\rm and} \quad ^{(1)}Z^{\alpha\beta}\neq 0\, .
\end{equation}
It has been verified by using our Reduce-Excalc computer algebra
programs that these are the only nonvanishing irreducible pieces of
the curvature. We find, in particular, $^{(5)}Z^{\alpha\beta}=0$.
Moreover, the strain curvature can be written in a compact notation as
\begin{equation}\label{1Z}
  ^{(1)}Z^{\alpha\beta} = \frac{1}{2V}d\left(\hat{\Phi}V^2\right)
  {\delta}^{\alpha}_{3}{\delta}^{\beta}_{3}\,\wedge{\ell} .
\end{equation}

The partial differential equation (\ref{dgl_V}) has simple polynomial
solutions,  inter alia, such as
\begin{equation}\label{solution_V}
V=f_{1}(u){\oz}\,^2 + f_{2}(u){\oz} + f_{3}(u) \qquad {\rm
or}\qquad V=f_{4}(u){\zeta}\,^2 + f_{5}(u){\zeta} + f_{6}(u)\,,
\end{equation}
with arbitrary wave profiles $f_{1}(u),\cdots\,,f_{6}(u)$.

Summarizing, the propagating {\em massless} spin 3-field can be
characterized, besides the coframe (\ref{coframe_0}), by
\begin{equation}
  ^{(1)}Q^{\alpha\beta}={\hat{\Phi}}V({\zeta}\, , {\oz}\, ,
  u){\delta}^{\alpha}_{3}{\delta}^{\beta}_{3}\, {\ell}=\hat{\Phi}V^2 
{\delta}^{\alpha}_{3}{\delta}^{\beta}_{3}\,\vartheta^2\,,
\end{equation}
where ${\hat\Phi}$ and $V$ are a solution of (\ref{dgl_V}). 
A comparison with (\ref{1Z}) shows that 
\begin{equation}
 ^{(1)}Z^{\alpha\beta}=\frac 12\,d^{(1)}Q^{\alpha\beta}\,,
\end{equation}
that is, the nonmetricity $^{(1)}Q^{\alpha\beta}$ acts as a true
potential for the strain curvature $ ^{(1)}Z^{\alpha\beta}$. The only
non zero component of the spin 3 carrying piece
$^{(1)}Q_{\alpha\beta}$ turns out to be
\begin{equation}\label{Q222}
^{(1)}Q_{222}={\hat{\Phi}}V^2(\zeta\, ,{\oz}\, , u)\,.
\end{equation}
Hence, the second field (\ref{Z_field2}) equation can be written
symbolically as
\begin{equation}
z_{1}\,\square \,\, ^{(1)}Q_{\alpha\beta\gamma} = 0\,.
\end{equation}

We would like to mention that all results in this subsection will
remain valid if one allows also for a non zero torsion trace, in
accordance with the general results of Heinicke et al.\ \cite{aether}.
Hence, any torsion trace could be parametrized as
\begin{equation}
^{(2)}T^{\alpha}={\Psi}{\vartheta}^{\alpha}\wedge {\ell}\, , \qquad
\Psi={\Psi}(\zeta\, ,\oz\, , u\, , v)\,,
\end{equation}
which is directly related to (\ref{ansatz_Q1}).

\subsubsection{Solutions with ${\ell}^2\neq 0$}\label{YMmassive}

In order to look for solutions of {\it massive propagating}
$^{(1)}Q_{\alpha\beta}$, we have to choose a more general
representation of the one-form ${\ell}$, because (\ref{ell}) describes
a null vector. As a simple modification of (\ref{ell}) leading to
non-vanishing ${\ell}^2$ we can choose
\begin{equation}\label{ell2_nonzero}
{\ell} = V{\vartheta}^{2} +
m_{0}{\vartheta}^{0}+m_{1}{\vartheta}^{1}\, ,
\end{equation}
where we assume for simplicity that $m_{0}$ and $m_{1}$ are
constants. For the norm ${\ell}^{2}$ we find
\begin{equation}\label{ell_2}
{\ell}^{2}=2m_{0}m_{1}\neq 0 \, .
\end{equation}
We could scale ${\ell}^2$ to unity with the choice
$m_{0}=m_{1}=1/{\sqrt{2}}$. However, we won't do so.

The ansatz (\ref{ans}) for the nonmetricity will be written
slightly modified as
\begin{equation}\label{Q_ell}
  Q^{\alpha\beta}=\frac{{\hat{\Phi}}{\ell}^{\alpha}{\ell}^{\beta}}{V{\ell}^2}
  \,{\ell}\,,
\end{equation}
with ${\hat{\Phi}}={\hat{\Phi}}(u)$ and ${\ell}^{\alpha}= e^{\alpha}
\rfloor {\ell}$. Even with these assumptions, it will be difficult to
solve the field equations. For this reason, we assume furthermore that
the scalar $V$ is constant, too.  This will lead us to a certain
toy-model showing that the solution manifold for the field equations
(\ref{Z_field}) and (\ref{Z_field2}) is not empty and allows for
massive propagating modes. We inserting all this into the first and
second field equation: The first field equation is fulfilled
identically, provided the coupling constants are chosen according to
\begin{eqnarray}\label{param_z}
5z_{1}+z_{3}+3(z_{4}+z_{5}) & = & 0\, , \\
5z_{1}+2z_{4}+z_{5} & = & 0\, ,
\end{eqnarray}
and the second field equation yields a second order linear
differential equation for $\hat{\Phi}$,
\begin{equation}\label{second_Phi}
(3z_{1}+z_{4})\hat{\Phi}_{uu} = 0\,.
\end{equation}
This simple model implies two different subcases, either
\begin{eqnarray} 
&&\hat{\Phi}(u) \quad {\rm arbitrary}\,,\quad{\rm
    with}\quad z_{3}=z_{5}=z_{1}\quad{\rm and} \quad z_{4}=-3z_{1}\, ,
  \quad {\and}\\
{\rm or}&&\nonumber \\ && {\hat{\Phi}}_{uu}=0 \, , \quad {\rm with}\quad
 5z_{1}+z_{3}+3(z_{4}+z_{5})  =  0 \quad {\rm and} \\ \nonumber
&&\hspace{70pt}5z_{1}+2z_{4}+z_{5} = 0\, , 
\end{eqnarray}
leading to a 2-parameter class of solutions.

We find for these solutions that only the strain-curvature
$Z_{\alpha\beta}$ is non-vanishing and that the nonmetricity
$Q_{\alpha\beta}$ is mainly non-trivial, namely
\begin{equation}\label{cond_curv}
  W_{\alpha\beta}=0\, , \quad ^{(2)}Z_{\alpha\beta}=0\, , \quad
  ^{(2)}Q_{\alpha\beta}=0\, .
\end{equation}
All other irreducible pieces are non-vanishing.

To give an idea of the complexity of this simple toy model, we list
the {\it massive} spin-3 part of the nonmetricity,
\begin{eqnarray}
  ^{(1)}Q_{00} & = & \frac{m_{1}\hat{{\Phi}}}{6m_{0}V}\left[
    m_{0}{\vartheta}^{0}+3\left(m_{1}{\vartheta}^{1}
      +V{\vartheta}^{2}\right)\right]\, , \\
 ^{(1)}Q_{01} & = &
  \frac{\hat{{\Phi}}}{6V}\left(m_{0}{\vartheta}^{0}
    +m_{1}{\vartheta}^{1}+2V{\vartheta}^{2}\right)\, , \\
  ^{(1)}Q_{02} & = & -\frac{m_{1}\hat{{\Phi}}}{6V}{\vartheta}^{2}\, ,
  \\ 
^{(1)}Q_{03} & = & -\frac{\hat{{\Phi}}}{6m_{0}V}\left[
    2m_{0}V{\vartheta}^{0}+3\left(m_{1}{\vartheta}^{1}+V{\vartheta}^{2}\right)V+
    m_{0}m_{1}{\vartheta}^{3}\right]\, , \\
^{(1)}Q_{11} & = & \frac{m_{0}\hat{{\Phi}}}{6m_{1}V}\left(
    3m_{0}{\vartheta}^{0}+m_{1}{\vartheta}^{1}+3V{\vartheta}^2\right)\, , \\
^{(1)}Q_{12} & = &
  -\frac{m_{0}\hat{{\Phi}}}{6V}{\vartheta}^{2}\,,\\ 
 ^{(1)}Q_{13} & =
  & -\frac{\hat{{\Phi}}}{6m_{1}V}\left[
    \left(3m_{0}{\vartheta}^{0}+2m_{1}{\vartheta}^{1}+3V{\vartheta}^{2}\right)V
    +m_{0}m_{1}{\vartheta}^{3}\right]\, , \\ 
 ^{(1)}Q_{22} & =
  & 0\, , \\ 
^{(1)}Q_{23} & = & \frac{\hat{{\Phi}}}{6V}\left(
    m_{0}{\vartheta}^{0}+m_{1}{\vartheta}^{1}+2V{\vartheta}^{2}\right)\, , \\
 ^{(1)}Q_{33} & = & \frac{\hat{{\Phi}}}{6m_{0}m_{1}}\left[
    3\left(m_{0}{\vartheta}^{0}+m_{1}{\vartheta}^{1}\right)V+3V^2{\vartheta}^2
    +2m_{0}m_{1}{\vartheta}^{3}\right]\, .
\end{eqnarray}
Because of (\ref{cond_curv}), also the spin-2 and spin-1
carrying pieces are non-trivial for those massive modes.
A systematic exploitation of the ansatz (\ref{ell2_nonzero}) and its
generalizations will be given elsewhere.

\subsubsection{Rewriting the Lagrangian $V_{\rm Z^2}$} \label{rewriting}

It is also instructive, to rewrite the Lagrangian (\ref{V_B}) in terms
of a set of different variables. The $^{(1)}Z_{\a\b}$ square piece we
leave as it is. Under our constraints, $^{(2)}Z_{\a\b}=0$, see
(\ref{result2}). The $^{(3)}Z_{\a\b}$, as displayed in
(\ref{result2}), can be expressed in terms of $d\Lambda$. This implies
\begin{equation}\label{Z3_sq}
^{(3)}Z_{\alpha\beta}\wedge\,
^{\star(3)}Z^{\alpha\beta}=\frac{1}{27}d{\Lambda}\wedge \,
^{\star}d{\Lambda}\, .
\end{equation}
Also simple is $^{(4)}Z_{\alpha\beta}$, see (\ref{Z4de}). Thus,
\begin{equation}\label{Z4_sq}
^{(4)}Z_{\alpha\beta}\wedge\,
^{\star(4)}Z^{\alpha\beta}=dQ\wedge{}^\star dQ\, .
\end{equation}
With the definition (\ref{Z5de}) of $^{(5)}Z_{\alpha\beta}$, we derive
the identity
\begin{equation}\label{Z5_sq}
^{(5)}Z_{\alpha\beta}\wedge\,
^{\star(5)}Z^{\alpha\beta}=\frac{3}{8}{\Xi}_{\alpha}\wedge\,
^{\star}{\Xi}^{\alpha}\, .
\end{equation}

Collecting our results (\ref{Z3_sq}), (\ref{Z4_sq}), (\ref{Z5_sq}),
and recalling $\Lambda=3Q$, see (\ref{QLam}), the Lagrangian
(\ref{V_B}) can be put into the form
\begin{equation}\label{V_B2}
  V_{\rm Z^2}=-\frac{1}{2\rho}\left[ z_{1}\,^{(1)}Z_{\alpha\beta}\wedge\,
    ^{\star(1)}Z^{\alpha\beta} +(\frac{z_3}{3}+z_4)dQ\wedge{}^\star dQ
    +\frac{3}{8}z_{5}{\Xi}_{\alpha}\wedge\,
    ^{\star}{\Xi}^{\alpha}\right]\, .
\end{equation}
If one desires, one can also introduce the Rizzi one-form. Under our
constraints, we have
\begin{equation}\label{V_B3}
  V_{\rm Z^2}=-\frac{1}{2\rho}\left[ z_{1}\,^{(1)}Z_{\alpha\beta}\wedge\,
    ^{\star(1)}Z^{\alpha\beta} +(\frac{z_3}{3}+z_4)dQ\wedge{}^\star dQ
    +\frac{3}{8}z_{5}
    {\riz}_{\alpha}\wedge\, ^{\star}{\riz}^{\alpha}\right]\, .
\end{equation}
Note that for a consistent transition to this new Lagrangian, one has
to add suitable Lagrange multiplier terms to the Lagrangian.

\section{Discussion}
\label{sec:discu}

In this paper, we carefully investigated the sector of MAG related to
a (free) massless spin-3 field and found exact solutions of full
nonlinear MAG theory in vacuum with propagating nonmetricity
$^{(1)}\!Q_{\a\b}\,$.

Up to an inessential factor $2$, we identified the spin-3 gauge field
in Vasiliev's frame formalism with $\qslash_{\a\b}\,$, the component
of the nonmetricity one-form which lies along the shear generator of
$GL(n,\mathbb{R})\subset \mathbb{R}^{n}\rtimes GL(n,\mathbb{R})\,$.
This enabled us to show in a direct way the appearance of Fronsdal's
massless spin-3 action in flat space as a part of MAG's action,
provided that the free parameters present in the latter action are
picked according to (\ref{valuesVa}), the remaining ones being zero
altogether.  Fronsdal's Lagrangian turns out to be purely quadratic in
the shear curvature, a purely post-Riemannian piece of the general
linear curvature.  We also clarified the dynamical spin content of the
plane-wave solution found in \cite{Yuripp-waves2006} by explicitly
relating it to a simple propagating helicity-3 solution of the
Bargmann--Wigner equations.

We then constructed several exact solutions of full nonlinear MAG in
vacuum with propagating tracefree nonmetricity, some showing a
massless spin-3 behavior, others presenting a massive-like spin-3
character.  Note that, although we have proved the occurrence of
Fronsdal's massless spin-3 Lagrangian inside MAG by choosing the only
nonzero parameters as in (\ref{valuesVa}), we have not shown that the
Singh--Hagen massive spin-3 Lagrangian \cite{Singh:1974qz} could also
be hosted inside MAG.  This would require the introduction of a scalar
field, not present in the general MAG Lagrangian (\ref{QMA}) we have
been considering here.  This scalar field was introduced in
\cite{Boulanger:2006tg} as a BEH field, in analogy to the Higgs field
in $U(1)$ symmetry breaking.

In MAG, as in any gauge theory, the geometrical fields are coupled to
matter currents. In addition to the symmetric (Hilbert)
energy-momentum current, which is coupled to the metric field, we have
additionally the spin current and the dilation plus shear currents
inducing torsion and nonmetricity fields, respectively.

This requires the homogeneous Lorentz group to be embedded in the
larger general linear group. Having identified Vasiliev's spin-3 frame
field with the traceless nonmetricity, we have gained another
geometrical interpretation for the former field (the tracelessness of
the Vasiliev gauge parameter $\hat{\xi}_{\a\b}$ being the natural
consequence of a shear transformation), but we have lost the Lorentz
group as the local symmetry group of the tangent manifold
\cite{Ne'eman:1996iz}.  Indeed, although the Weyl one-form leaves the
conformal light-cone structure intact, the traceless nonmetricity
(which couples to the shear current of matter) does {\it not} preserve
the light-cone structure and the local Lorentz symmetry under parallel
transport \cite{PRs} with respect to the connection $\Gamma_\a{}^\b$.
This implies that, in our discussion, we are relating the massless
spin-3 field with situations in which there is no conventional flat,
Special Relativity limit, like e.g. in the early universe or in the
microscopic domain where the coupling of the shear plus dilation
current of matter to nonmetricity is expected to become
non-negligible, not to mention the coupling of matter's intrinsic spin
current to the torsion field.
This picture is in accordance with Fronsdal's spin-3 Lagrangian inside
MAG being purely quadratic in the shear curvature, hence belonging to
the strong-gravity post-Riemannian part of MAG's Lagrangian.

Although there is presumably no consistent coupling between a spin-3
field and dynamical Hilbert--Einstein gravity (without resorting to an
infinite tower of higher-spin fields), our results suggest that spin-3
dynamics in the framework of MAG could be well-defined in the limit
where strong-gravitational MAG effects prevail and where shear-type
excitations of matter are expected to arise. Finally, it would be interesting
to compare our results with those presented in \cite{aether}.


\section*{Acknowledgements}

N.B.\ wants to thank Ingo Kirsch for discussions on the possible
presence of Fronsdal's massless spin-3 theory in MAG. He is also
grateful to the Gravitation and Relativity Group of K\"oln University
and, particularly, to Claus Kiefer for kind hospitality. The work of
N.B. was supported by the Fonds National de la Recherche Scientifique
(Belgium). P.B.\ and F.W.H.\ would like to thank Ingo Kirsch for
helpful remarks that he made to a draft version of our paper.

\appendix

\section{Irreducible decomposition of the strain
  curvature}\label{irrdec.Z}

\subsection{In components}

We have the following irreducible decomposition of the components of
the strain curvature two-form
$Z_{\alpha\beta}=\frac{1}{2}\,Z_{\g\d\alpha\beta}\,\vartheta^{\g}\wedge
\vartheta^{\d}=\frac{1}{2} Z_{ij\a\b}\,dx^i\wedge dx^j$ with respect to
the (pseudo)-orthogonal group, cf.\ \cite{PRs,aether,WeylMeet},
\begin{equation}\label{ZdecompYT}
\underbrace{Z_{\alpha\beta}}_{
{\begin{picture}(25,10)(0,0)
\multiframe(1,4)(4.5,0){1}(4,4){}
\multiframe(1,-.5)(4.5,0){1}(4,4){}
\put(7,4){\tiny{$\otimes$}}
\multiframe(15,4)(4.5,0){2}(4,4){}{}
\end{picture}
}}
=\underbrace{^{(1)}Z_{\alpha\beta}}_{
{\begin{picture}(15,10)(0,0)
\multiframe(1,4)(4.5,0){3}(4,4){}{}{}
\multiframe(1,-.5)(4.5,0){1}(4,4){}
\end{picture}
}}\,
\oplus \underbrace{^{(2)}Z_{\alpha\beta}}_{
{\begin{picture}(12,10)(0,0)
\multiframe(1,4)( 4.5,0){2}(4,4){}{}
\multiframe(1,-.5)(4.5,0){1}(4,4){}
\multiframe(1,-5)(4.5,0){1}(4,4){}
\end{picture}
}}\,
\oplus \underbrace{^{(3)}Z_{\alpha\beta}}_{
{\begin{picture}(8,10)(0,0)
\multiframe(1,4)(4.5,0){1}(4,4){}
\multiframe(1,-.5)(4.5,0){1}(4,4){}
\end{picture}
}}\,
\oplus \underbrace{^{(4)}Z_{\alpha\beta}}_{
{\begin{picture}(8,10)(0,0)
\multiframe(1,4)(4.5,0){1}(4,4){}
\multiframe(1,-.5)(4.5,0){1}(4,4){}
\end{picture}
}}\,
\oplus \underbrace{^{(5)}Z_{\alpha\beta}}_{
{\begin{picture}(8,10)(0,0)
\multiframe(1,4)(4.5,0){2}(4,4){}{}
\end{picture}
}}\quad \,.
\end{equation}
We have given the decomposition of the $GL(n,\mathbb R)$-reducible
components $Z_{\g\d\alpha\beta}$ into irreducible representations of
the (pseudo)-orthogonal group, so that the Young diagrams on the
right-hand-side of the above equality label $O(1,n-1)$-irreducible
representations. (Note the multiplicity 2 of the antisymmetric rank-2
tensor irreducible representation.  Indeed, $Z_{\a[\g\d]}^{~~~~\;\a}$
and $Z_{\g\d\a}^{~~~\;\a}$ are linearly independent.) Accordingly,
\begin{eqnarray}
  {}^{(1)}Z_{\g\d   \a\b} &=& 
  \frac{1}{2}\,(\zslash_{\g\d \a\b}-\zslash_{\a[\g \d]\b}-\zslash_{\b[\g \d]\a})
\nonumber \\
  && +\; \frac{1}{2(n+2)}\,\left(
    \zslash _{\varepsilon[\g \alpha]}^{\quad~\;\varepsilon}g_{\beta\d} -
    \zslash _{\varepsilon[\d \alpha]}^{\quad~\;\varepsilon}g_{\beta\g} +
    \zslash _{\varepsilon[\g \beta]}^{\quad~\;\varepsilon}g_{\alpha\d} -
    \zslash _{\varepsilon[\d \beta]}^{\quad~\;\varepsilon}g_{\alpha\g}
    + 2 g_{\alpha\beta}\zslash _{\varepsilon[\g \d]}^{\quad~\;\varepsilon}
  \right)
  \nonumber \\
  &&+\;\frac{1}{n}\,\left(
    \zslash _{\varepsilon(\g \alpha)}^{\quad~\;\varepsilon} g_{\beta\d}
    + \zslash _{\varepsilon(\g \beta)}^{\quad~\;\varepsilon} g_{\alpha\d}
    - \zslash _{\varepsilon(\d \alpha)}^{\quad~\;\varepsilon} g_{\beta\g}
    - \zslash _{\varepsilon(\d \beta)}^{\quad~\;\varepsilon} g_{\alpha\g}
  \right) \,,
  \nonumber \\
  {}^{(2)}Z_{\g\d   \a\b} &=&  
  \frac{1}{2(n-2)}\,\left(
    \zslash _{\varepsilon[\g \alpha]}^{\quad~\;\varepsilon} g_{\beta\d}
    + \zslash _{\varepsilon[\g \beta]}^{\quad~\;\varepsilon} g_{\alpha\d}
    - \zslash _{\varepsilon[\d \alpha]}^{\quad~\;\varepsilon} g_{\beta\g}
    - \zslash _{\varepsilon[\d \beta]}^{\quad~\;\varepsilon} g_{\alpha\g}
  \right)
  \nonumber \\
  &&+\; \frac{3}{4}\,
  (\zslash _{[\g\d \alpha]\beta}+\zslash _{[\g\d \beta]\alpha} )
  - \; \frac{1}{(n-2)}\,
  \zslash _{\varepsilon[\g \d]}^{\quad~\varepsilon}g_{\alpha\beta}\,,
  \nonumber \\
  {}^{(3)}Z_{\g\d   \a\b} &=&  \frac{n}{(n+2)(n-2)}\,\left(
    \zslash _{\varepsilon[\alpha \g]}^{\quad~\;\varepsilon} g_{\beta\d}
    + \zslash _{\varepsilon[\beta \g]}^{\quad~\;\varepsilon} g_{\alpha\d}
    - \zslash _{\varepsilon[\alpha \d]}^{\quad~\;\varepsilon} g_{\beta\g}
    - \zslash _{\varepsilon[\beta \d]}^{\quad~\;\varepsilon} g_{\alpha\g}
  \right)
  \nonumber \\
  &&+\;\frac{4}{(n+2)(n-2)}\,
  \zslash _{\varepsilon[\g \d]}^{\quad~\,\varepsilon}g_{\alpha\beta}\,,
 \nonumber \\
  {}^{(4)}Z_{\g\d   \a\b} &=&  \frac{1}{n}\,
  {Z}_{\g\d \varepsilon}^{\quad\,\varepsilon}g_{\alpha\beta}\,,
  \nonumber \\
  {}^{(5)}Z_{\g\d   \a\b} &=&  \frac{1}{n}\,
  \left(
    \zslash _{\varepsilon(\alpha \d)}^{\quad~\;\varepsilon} g_{\beta\g}
    + \zslash _{\varepsilon(\beta \d)}^{\quad~\;\varepsilon} g_{\alpha\g}
    - \zslash _{\varepsilon(\alpha \g)}^{\quad~\;\varepsilon} g_{\beta\d}
    - \zslash _{\varepsilon(\beta \g)}^{\quad~\;\varepsilon} g_{\alpha\d}
  \right)\,,
  \nonumber
\end{eqnarray}
and with the shear curvature
\begin{eqnarray}\label{Zslash}
\zslash _{\g\d \alpha\beta}:={Z}_{\g\d \alpha\beta}-\frac{1}{n}\,
g_{\alpha\beta}{Z}_{\g\d \varepsilon}^{\quad~\varepsilon}\,.\nonumber
\end{eqnarray}
Equivalently, by introducing $\Riz_{\a\b}$, we can rewrite this as
follows:
\begin{eqnarray}
^{(1)}Z_{\gamma\delta\alpha\beta} & = & \frac{1}{2} \left(
{\zslash}_{\gamma\delta\alpha\beta} -
{\zslash}_{\alpha[\gamma\delta]\beta}
-{\zslash}_{\beta[\gamma\delta]\alpha} \right)\nonumber\\
 & & 
+ \frac{1}{2(n+2)}\left( {\Riz}_{[\gamma\alpha
]}g_{\beta\delta}-{\Riz}_{[\delta\alpha ]}g_{\beta\gamma} +
{\Riz}_{[\gamma\beta ]}g_{\alpha\delta}-{\Riz}_{[\delta\beta
]}g_{\alpha\gamma}\right) \nonumber \\
 & & +\frac{1}{n+2}{\Riz}_{[\gamma\delta
]}g_{\alpha\beta}\nonumber\\
& & + \frac{1}{n}\left(
{\Riz}_{(\gamma\alpha )}g_{\beta\delta} + {\Riz}_{(\gamma\beta
)}g_{\alpha\delta} - {\Riz}_{(\delta\alpha )}g_{\beta\gamma}
-{\Riz}_{(\delta\beta )}g_{\alpha\gamma} \right)\nonumber \\
 ^{(2)}Z_{\gamma\delta\alpha\beta} & = &
\frac{1}{2(n-2)}\left[ \left( {\Riz}_{[\gamma |
\beta}-{\Riz}_{\beta [\gamma}\right)g_{|\alpha |\delta ]}+ \left(
{\Riz}_{[\gamma |\alpha}-{\Riz}_{\alpha |\gamma ]}\right)g_{\beta
| \delta ]}\right] \nonumber \\
 & &  + \, \frac{3}{4}\left(
{\zslash}_{[\gamma\delta\alpha
]\beta}+{\zslash}_{[\gamma\delta\beta ]\alpha}\right)-
\frac{1}{n-2}{\Riz}_{[\gamma\delta ]}g_{\alpha\beta}\, , \nonumber \\
 ^{(3)}Z_{\gamma\delta\alpha\beta} & = &
\frac{n}{n^2-4}\left[ \left( {\Riz}_{\alpha
[\gamma}-{\Riz}_{[\gamma |\alpha}\right)g_{|\beta |\delta ]} +
\left( {\Riz}_{\beta [\gamma}-{\Riz}_{[\gamma
|\beta}\right)g_{|\alpha |\delta ]}\right] \nonumber \\
& & + \, \frac{4}{n^2-4}{\Riz}_{[\gamma\delta
  ]}g_{\alpha\beta}\, , \nonumber \\
 ^{(4)}Z_{\gamma\delta\alpha\beta} & = &
\frac{1}{n}Z_{\gamma\delta\rho}{}^{\rho}g_{\alpha\beta}\, ,
\nonumber \\
 ^{(5)}Z_{\gamma\delta\alpha\beta} & = &
\frac{1}{n}\left[ \left( {\Riz}_{\alpha [\delta} + {\Riz}_{[\delta
      |\alpha |}\right)g_{\gamma ]\beta} + \left( {\Riz}_{\beta [\delta}
    + {\Riz}_{[\delta |\beta |}\right)g_{\gamma ]\alpha}\right]\,
.\nonumber
\end{eqnarray}

With regard to the uniqueness of the decomposition, a remark is in
order: If we simply apply the Young diagram procedure to the
components $Z_{\g\d\a\b}$ of $Z_{\a\b}$ and take traces, three of
the five irreducible pieces obtained are $^{(1)}\!Z_{\a\b}$,
$^{(2)}\!Z_{\a\b}\,$, and $^{(5)}\!Z_{\a\b}$, as above, but the
remaining two pieces are arbitrary combinations of the two irreducible
subspaces involved in $^{(3)}\!Z_{\a\b}$ and $^{(4)}\!Z_{\a\b}$ above
and hence are not canonical. Here, however, the initial decomposition
(\ref{Zslash}) with respect to the indices on the {\it two-form\/}
$Z_{\a\b}$ has led to a unique canonical set of irreducible pieces.

\subsection{In exterior calculus, analogies with the irreducible
  decomposition of $Q_{\a\b}$}

We recall the definition of the tracefree shear curvature two-form
\begin{equation}\label{zslashdef}
  \zslash_{\a\b}=Z_{\a\b}-\frac{1}{n}g_{\a\b}Z\,,\quad{\rm with}
\quad Z=Z_\g{}^\g\,.
\end{equation}
We cut this two-form into different pieces by contracting with $e_\b$
and transvecting with $\vta^\a$:
\begin{equation}\label{ZDY}\zslash_\a:=e^\b\rfloor
  \zslash_{\a\b}\equiv \Riz_\a\,, \quad
  \hat\Delta:={1\over n-2}\,\vta^\a\wedge\zslash_\a\,,\quad
S_\a:=\zslash_{\a\b}\wedge\vt^\b+\vt_\a\wedge\hat{\Delta}\,.
\end{equation}
We have $\vt^\a\wedge S_\a=0$, $e^\a\rfloor S_\a=0$, that is, the
three-form $S_\a$, in 4D, has $4\times 4-1-6=9$ independent
components.  Subsequently we can subtract out the trace of
$\zslash_\a$:
\begin{equation}\Xi_\a:=
  \zslash_\a-{1\over2}e_\a\rfloor(\vta^\g\wedge\zslash_\g)\,.
\end{equation}
We have $\vt^\a\wedge \Xi_\a=0$, $e^\a\rfloor \Xi_\a=0$, that is, the
one-form $\Xi_\a$, in 4D, has $4\times 4-6-1=9$ independent
components.

The irreducible pieces may then be written as (the number of
independent components is specified for $n=4$)
\begin{eqnarray}
(9\;{\rm ind.comp.})\qquad
  ^{(2)}\!Z_{\a\b}&:=& {1\over 2}\,e_{(\a}\rfloor S_{\b)}\,,\\
(6\;{\rm ind.comp.})\qquad
\label{Z3de}^{(3)}\!Z_{\a\b}&:=&{n\over n+2}\;\left(
  \vta_{(\a}\wedge e_{\b)}\rfloor-\frac{2}{n}\,g_{\a\b}\right)\hat\Delta\,,\\
(6\;{\rm ind.comp.})\qquad
\label{Z4de}^{(4)}\! Z_{\a\b}&:=& {1\over n}\;g_{\a\b}\,Z\,,\\
(9\;{\rm ind.comp.})\qquad
\label{Z5de}^{(5)}\! Z_{\a\b}&:=&{2\over n}\; \vta_{(\a}\wedge
\Xi_{\b)}\,,\\
(30\;{\rm ind.comp.})\qquad^{(1)}\!Z_{\a\b}&:=& Z_{\a\b}-
\,^{(2)}\!Z_{\a\b}-\,^{(3)}\!Z_{\a\b}-\,^{(4)}\!Z_{\a\b}-
\,^{(5)}\!Z_{\a\b}\,.\end{eqnarray}
Apparently, the forms $\{S_\a,\,\hat{\Delta},\,Z,\,\Xi_\a\}$ are
equivalent to the irreducible pieces $
\{^{(2)}\!Z_{\a\b},{}^{(3)}\!Z_{\a\b},{}^{(4)}\! Z_{\a\b},{}^{(5)}\!
Z_{\a\b}\}$, 
respectively. 

The strain curvature is of the type of a field strength. The
corresponding ``potential'' is expected to be the nonmetricity
$Q_{\a\b}$. As we will show, the irreducible decomposition of the
nonmetricity is reminiscent of the that of the strain curvature. In
order to underline this, we will present all definitions etc.\
strictly in parallel to the formulas above of the strain curvature.

We start with the tracefree nonmetricity one-form
\begin{equation}\label{qslashdef}
  \qslash_{\a\b}=Q_{\a\b}-g_{\a\b}Q\,,\quad{\rm with}
  \quad Q=\frac{1}{n}Q_\g{}^\g\,.
\end{equation}
We cut this two-form into different pieces by contracting with $e_\b$
and transvecting with $\vta^\a$:
\begin{equation}\label{Lambda}
  \Lambda_\a := e^\beta \rfloor\!\qslash_{\alpha\beta}\,,\quad
\Lambda:=\vt^\a\Lambda_\a\,,\quad P_\a:=\qslash_{\a\b}\wedge
\vt^\b-\frac{1}{n-1}\vt_\a\wedge\Lambda\,.
\end{equation}
We have $\vt^\a\wedge P_\a=0$, $e^\a\rfloor P_\a=0$, that is, the
two-form $P_\a$, in 4D, has $6\times 4-4-4=16$ independent
components.

The irreducible pieces may then be written as (the number of
independent components is specified for $n=4$)
\begin{eqnarray}\label{deco4}
  (16\;{\rm ind.comp.})\qquad  
  {}^{(2)}Q_{\alpha\beta} & := & -\frac{2}{3} \, e_{(\alpha} \rfloor
  P _{\beta)}\label{Q2} \,,\\
  (4\;{\rm ind.comp.})\qquad
  {}^{(3)}Q_{\alpha\beta} & := & \frac{2n}{(n-1)(n-2)} \left(
    \vartheta_{(\alpha} \,e_{\beta )}\rfloor - \frac{1}{n} \,
    g_{\alpha\beta} \right)\Lambda \,,\label{3Qa} \\
  (4\;{\rm ind.comp.})\qquad
  {}^{(4)}Q_{\alpha\beta} & := & Q \, g_{\alpha\beta} \,, \\
  (16\;{\rm ind.comp.})\qquad  {}^{(1)}Q_{\alpha\beta} &:=& Q_{\alpha\beta} -
  {}^{(2)}Q_{\alpha\beta} - {}^{(3)}Q_{\alpha\beta} -
  {}^{(4)}Q_{\alpha\beta} \,.\label{deco1}
\end{eqnarray}
Apparently, the forms $\{P_\a,\,\Lambda,\,Q\}$ are equivalent to the
irreducible pieces $ \{^{(2)}\!Q_{\a\b},\, {}^{(3)}\!Q_{\a\b},\,
{}^{(4)}\!  Q_{\a\b}\}$, respectively.

The analogies between the different irreducible decomposition of the
forms $T^{\alpha}$, $Q^{\alpha\beta}$, and $Z^{\alpha\beta}$ in $n$
dimensions can be displayed in a pictorial description as follows:
\begin{eqnarray}
  \pmatrix{T\cr \mathfrak{A}} \quad \lapl \, \quad\hspace{7pt}
  T^{\alpha} &= &\pmatrix{
    ^{(1)}T^{\alpha}\cr ^{(2)}T^{\alpha}\cr ^{(3)}T^{\alpha} }\\
  \pmatrix{ P_{\alpha}\cr {\Lambda}\cr Q} \quad \lapl \, \quad Q_{\alpha\beta}& =&
  \pmatrix{^{(1)}Q^{\alpha\beta}\cr ^{(2)}Q^{\alpha\beta}\cr
    ^{(3)}Q^{\alpha\beta}\cr ^{(4)}Q^{\alpha\beta}}\\
\pmatrix{ S_{\alpha}\cr {\hat{\Delta}}\cr Z\cr {\Xi}_{\alpha}} \quad
\lapl \, \quad
 Z_{\alpha\beta}& =& \pmatrix{^{(1)}Z^{\alpha\beta}\cr ^{(2)}Z^{\alpha\beta}\cr
 ^{(3)}Z^{\alpha\beta}\cr ^{(4)}Z^{\alpha\beta}\cr ^{(5)}Z^{\alpha\beta} }
\end{eqnarray}
where the symbol $\lapl$ denotes the correspondence between the set of
forms on the left-hand-side and the corresponding irreducible pieces
of the field strengths on the right-hand-side. Hence, the common
procedure shows that we need $k$ independent forms (generally of
different degrees) to create $k+1$ irreducible pieces of the
corresponding field strength. We recall the definition
$T:=e_{\alpha}\rfloor\, T^{\alpha}$ and of $^{(3)}T^\a:=
e^\a\rfloor\mathfrak{A}$, together with
$\mathfrak{A}:=\frac{1}{3}\vt^\b\wedge T_\b$.

For later convenience, we list the irreducible pieces as wedged with
$\vt^\b$:
\begin{eqnarray}\label{Zwedge}
{}^{(1)}Z_{\a\b}\wedge\vt^\b &=&0\,,\nonumber\\
{}^{(2)}Z_{\a\b}\wedge\vt^\b &=&S_\a\,,\nonumber\\
{}^{(3)}Z_{\a\b}\wedge\vt^\b &=&-\vt_\a\wedge\hat{\Delta}\,,\nonumber\\
{}^{(4)}Z_{\a\b}\wedge\vt^\b &=&\frac{1}{n}\vt_\a\wedge Z\,,\nonumber\\
{}^{(5)}Z_{\a\b}\wedge\vt^\b &=&0\,,\nonumber\\
\zslash_{\a\b}\wedge\vt^\b &=&S_\a-\vt_\a\wedge\hat{\Delta}\,.
\end{eqnarray}
We can do the analogous for the nonmetricity:
\begin{eqnarray}\label{Qwedge}
{}^{(1)}Q_{\a\b}\wedge\vt^\b &=&0\,,\nonumber\\
{}^{(2)}Q_{\a\b}\wedge\vt^\b &=&P_\a\,,\nonumber\\
{}^{(3)}Q_{\a\b}\wedge\vt^\b &=&\frac{1}{n-1}\vt_\a\wedge\Lambda\,,\nonumber\\
{}^{(4)}Q_{\a\b}\wedge\vt^\b &=&-\vt_\a\wedge Q\,,\nonumber\\
\qslash_{\a\b}\wedge\vt^\b &=&P_\a+\frac{1}{n-1}\vt_\a\wedge\Lambda\,.
\end{eqnarray}

\section{Zeroth Bianchi identity}\label{zeroB}

\subsection{Zeroth Bianchi
  identity in different disguises}

A link between the three-form $S_\a\sim{}^{(2)}Z_{\a\b}$ and the
two-form $P_\a\sim {}^{(2)}Q_{\a\b}$ can be found via the zeroth
Bianchi identity:
\begin{equation}\label{0thBia}
DQ_{\a\b}\equiv 2Z_{\a\b}\,.
\end{equation}
We introduce the slashed quantities:
\begin{equation}\label{0thBia1}
  D\qslash_{\a\b}+D(Qg_{\a\b})=2\zslash_{\a\b}+\frac{2}{n}g_{\a\b}
Z_\g{}^\g
\end{equation}
or, since $dQ=\frac{2}{n}Z_\g{}^\g$,
\begin{equation}\label{0thBia2}
 D\qslash_{\a\b}+dQg_{\a\b}+Q\wedge Q_{\a\b}=2\zslash_{\a\b}+g_{\a\b}dQ\,.
\end{equation}
Accordingly,
\begin{equation}\label{0thBia3}
D\qslash_{\a\b}+Q\wedge \qslash_{\a\b}=2\zslash_{\a\b}\,.
\end{equation}

The difference between the connection $\Gamma_\a{}^\b$ and the
Riemannian connection $\widetilde{\Gamma}_\a{}^\b$ is the distortion
one-form
\begin{equation}\label{distortion}
  N_\a{}^\b=\Gamma_\a{}^\b-\widetilde{\Gamma}_\a{}^\b\,,\quad{\rm with}
\quad N_{(\a\b)}=\frac{1}{2}Q_{\a\b}\,,\; N_\b{}^\a\wedge\vt^\b=T^\a\,.
\end{equation}
If we execute the covariant exterior differentiation in
(\ref{0thBia3}), we find
\begin{eqnarray}\label{0thBia4}
  &&\widetilde{D}\qslash_{\a\b}-N_{[\a\g]}\wedge\qslash^{\;\g}{}_\b
  -N_{[\b\g]}\wedge\qslash_\a{}^\g
  +Q\wedge \qslash_{\a\b}\nonumber\\
  &&\hspace{40pt}-\frac 12Q_{\a\g}\wedge\qslash^{\;\g}{}_\b
  -\frac 12Q_{\b\g}\wedge\qslash_\a{}^\g
  =2\zslash_{\a\b}\,.\end{eqnarray}
After some algebra, the explicit square pieces in the nonmetricity drop
out. Thus,
\begin{eqnarray}\label{0thBia5}
  &&\widetilde{D}\qslash_{\a\b}-N_{[\a\g]}\wedge\qslash^{\;\g}{}_\b
  -N_{[\b\g]}\wedge\qslash_\a{}^\g
 =2\zslash_{\a\b}\,.
\end{eqnarray}

Let us come back to (\ref{0thBia3}). We wedge {}from the
right-hand-side with $\vt^\b$:
\begin{equation}\label{0thBia4a}
  D(\qslash_{\a\b}\wedge \vt^\b)
  +Q\wedge( \qslash_{\a\b}\wedge\vt^\b)+\qslash_{\a\b}\wedge T^\b
  =2(\zslash_{\a\b}\wedge\vt^\b)\,.
\end{equation}
Also here we can provide a version with a Riemannian derivative. The
simplest is to wedge (\ref{0thBia5}) from the right with $\vt^\b$ and
to note $\widetilde{D}\vt^\a=0$:
\begin{eqnarray}\label{riem}
  &&(\widetilde{D}\qslash_{\a\b}\wedge\vt^\b)-N_{[\a\g]}\wedge
\qslash^{\;\g}{}_\b\wedge\vt^\b
  -N_{[\b\g]}\wedge\qslash_\a{}^\g\wedge\vt^\b
 =2\zslash_{\a\b}\wedge\vt^\b\,.
\end{eqnarray}

Then we substitute (\ref{Qwedge}) and (\ref{Zwedge}) into
(\ref{0thBia4a}) and find
\begin{equation}\label{0thBia5a}
  D(P_\a+\frac{1}{n-1}\,\vt_\a\wedge\Lambda)
  +Q\wedge(P_\a+\frac{1}{n-1}\,\vt_\a\wedge\Lambda)+\qslash_{\a\b}\wedge T^\b
  =2(S_\a-\vt_\a\wedge\hat{\Delta})\,.
\end{equation}
We differentiate the sum and collect the torsion dependent terms
\begin{equation}\label{0thBia6}
 ( DP_\a+Q\wedge P_\a)-\frac{1}{n-1}\,\vt_\a\wedge(d\
  \Lambda+Q\wedge \Lambda)
  +(\qslash_{\a\b}+\frac{1}{n-1}\,g_{\a\b}\Lambda)\wedge T^\b
  =2(S_\a-\vt_\a\wedge\hat{\Delta})\,.
\end{equation}

Our strategy is now to separate $S_\a$ from $\hat{\Delta}$. We
contract (\ref{0thBia6}) from the left with $-\frac 14 e^\a\rfloor$:
\begin{equation}\label{0thBia7}
  -\frac 14 e^\g\rfloor\{{\rm l.h.s.of\;}(\ref{0thBia6})\}_\g=
\frac 12 e^\a\rfloor(\vt_\a\wedge\hat{\Delta})=2\hat{\Delta}
-\frac 12\vt_\a\wedge(e^\a\rfloor\hat{\Delta})=\hat{\Delta}
\end{equation}
or
\begin{equation}\label{0thBia8}
  ^{(3)}Z_{\a\b}\sim\quad\hat{\Delta}= -\frac 14 e^\g\rfloor\{{\rm
    l.h.s.of\;}  (\ref{0thBia6})\}_\g\,.
\end{equation}
Now we can resolve (\ref{0thBia6}) with respect to $S_\a$:
\begin{eqnarray}\label{0thBia9}
  ^{(2)}Z_{\a\b}\sim\quad  S_\a&=&\frac 12\left( DP_\a+
    Q\wedge P_\a-\frac{1}{n-1}\,\vt_\a\wedge(d\
    \Lambda+Q\wedge \Lambda)
   \right.\nonumber\\
&&\left.\hspace{25pt} +(\qslash_{\a\b}+\frac{1}{n-1}\,g_{\a\b}\Lambda)\wedge
    T^\b\right)+\vt_\a\wedge\hat{\Delta}\,.
\end{eqnarray}
In this formula, ${}^{(2)}Z_{\a\b}\sim S_\a$ is expressed in terms of
nonmetricity and torsion. Note that our results (\ref{0thBia9}) and
(\ref{0thBia8}) are generally valid. No constraints have been assumed
so far. However, this will be done in the next subsection.

\subsection{Consequences of the ansatz (\ref{ans}) and of
  vanishing torsion (\ref{vanT})}

We substitute (\ref{QLam}), (\ref{P2Q}), and (\ref{vanT}) into
(\ref{0thBia9}):
\begin{equation}\label{0thBia10}
  S_\a=-\frac{1}{2(n-1)}\,\vt_\a\wedge d\
  \Lambda+\vt_\a\wedge\hat{\Delta}\,.
\end{equation}
The two-form $\hat{\Delta}$ we take from (\ref{0thBia8}) after the
constraints (\ref{QLam}), (\ref{P2Q}), and (\ref{vanT}) have been
substituted into the left-hand-side of (\ref{0thBia6}). Thus,
\begin{equation}\label{result1}
 \hat{\Delta}=\frac{1}{2(n-1)} d \Lambda
 \qquad{\rm and}\qquad S_\a=0
\end{equation}
or 
\begin{equation}\label{result2}
  ^{(2)}\!Z_{\a\b}=0 \,\qquad{\rm and}\qquad{}^{(3)}\!Z_{\a\b}=
  \frac{n}{2(n-1)(n+2)}\left(
    \vta_{(\a}\wedge e_{\b)}\rfloor-\frac{2}{n}
    \,g_{\a\b}\right) d \Lambda\,.
\end{equation}

\section{First Bianchi identity}\label{firstB}

Consider the first Bianchi identity,
\begin{eqnarray}\label{first_Bianchi}
DT^{\alpha} & \equiv &  R_{\b}{}^{\alpha}\wedge {\vartheta}^{\b}\,.
\end{eqnarray}
The irreducible pieces of $W_{\a\b}$ and $Z_{\a\b}$ obey quite
generally the algebraic constraints \cite{PRs}
\begin{eqnarray}
{}^{(1)}W_{\b}{}^{\alpha}\wedge\vt^\b&=&
{} ^{(4)}W_{\b}{}^{\alpha}\wedge\vt^\b=
{} ^{(6)}W_{\b}{}^{\alpha}\wedge\vt^\b\nonumber\\
&=&{} ^{(1)}Z_{\b}{}^{\alpha}\wedge\vt^\b=
{} ^{(5)}Z_{\b}{}^{\alpha}\wedge\vt^\b=0\,.
\end{eqnarray}
Thus,
\begin{eqnarray}
  DT^{\alpha} = \left(
    ^{(2)}W_{\b}{}^{\alpha} + \, ^{(3)}W_{\b}{}^{\alpha} + \,
    ^{(5)}W_{\b}{}^{\alpha} + \, ^{(2)}Z_{\b}{}^{\alpha} + \,
    ^{(3)}Z_{\b}{}^{\alpha} + \, ^{(4)}Z_{\b}{}^{\alpha}\right) \wedge
  {\vartheta}^{\b} = 0\, .
\end{eqnarray}

\section{Second Bianchi identity}\label{secondB}

The second Bianchi identity reads
\begin{eqnarray}\label{2ndB}
  DR_\b{}^\a&=&D\left(Z_\b{}^\a+W_\b{}^\a \right)
  =D\left(\sum\limits_{I=1}^{5}{}^{(I)}Z_\b{}^\a+
    \sum\limits_{I=1}^{6}{}^{(I)}W_\b{}^\a \right)\cr
  &=&D\left(\sum\limits_{I=1}^{5}({}^{(I)}Z_\b{}^\a+
    {}^{(I)}W_\b{}^\a)-\frac{W}{12}\vt_\b\wedge\vt^\a \right)\equiv 0\,.
\end{eqnarray}
Here $W:=e_\g\rfloor e_\d\rfloor W^{\g\d}$ is the curvature scalar and
the corresponding term in (\ref{2ndB}) represents the sixth piece of
$W_\b{}^\a$, see \cite{PRs}.


\centerline{=======}
\end{document}